\documentclass[%
 preprint,
superscriptaddress,
 amsmath,amssymb,
 aps,
]{revtex4-1}

\usepackage{graphicx}
\usepackage{dcolumn}
\usepackage{bm}
\usepackage{verbatim}
\usepackage{color,soul}
\usepackage{epstopdf}
\AppendGraphicsExtensions{.tif}

\begin{document}


\title{Thickness and temperature dependence of the atomic-scale structure of SrRuO$_3$ thin films}

\author{Xuanyi Zhang}
 \affiliation{Department of Physics, North Carolina State University. Raleigh, NC, 27695, USA}
 
 \author{Aubrey N. Penn}
\affiliation{Department of Materials Science and Engineering, North Carolina State University. Raleigh, NC, 27695, USA}

 \author{Lena Wysocki}
  \affiliation{University of Cologne, Institute of Physics II, 50937 Cologne, Germany}
  
  \author{Zhan Zhang}
\affiliation{Advanced Photon Source, Lemont, IL 76019, USA}

  \author{Paul H. M. van Loosdrecht}
  \affiliation{University of Cologne, Institute of Physics II, 50937 Cologne, Germany}

   \author{Lior Kornblum}
  \affiliation{Andrew \& Erna Viterbi Department of Electrical Engineering, Technion Israel Institute of Technology, 3200003 Haifa, Israel}
 
  \author{James M. LeBeau}
  \affiliation{Department of Materials Science and Engineering, Massachusetts Institute of Technology, Cambridge, MA 02139, USA}
  
 \author{Ionela Lindfors-Vrejoiu}
 \email{vrejoiu@ph2.uni-koeln.de}
  \affiliation{University of Cologne, Institute of Physics II, 50937 Cologne, Germany}
  
\author{Divine P. Kumah}
\email{dpkumah@ncsu.edu}
 \affiliation{Department of Physics, North Carolina State University. Raleigh, NC, 27695, USA}  

\date{\today}

\begin{abstract}
Due to the strong lattice-property relationships which exist in complex oxide epitaxial layers, their electronic and magnetic properties can be modulated by structural distortions induced at the atomic scale. The modification and control can be affected at coherent heterointerfaces by epitaxial strain imposed by the substrate or by structural modifications to accommodate the film-substrate symmetry mismatch. Often these act in conjunction with a strong dependence on the layer thickness, especially for ultrathin layers. Moreover, as a result of these effects, the temperature dependence of the structure may deviate largely from that of the bulk. The temperature-dependent structure  of 3 to 44 unit cell thick ferromagnetic SrRuO$_3$ films grown on Nb-doped SrTiO$_3$ substrates are investigated using a combination of high-resolution synchrotron X-ray diffraction and high resolution electron microscopy. This aims to shed light on the intriguing magnetic and  magnetotransport properties of epitaxial SRO layers, subjected to extensive investigations lately. The oxygen octahedral tilts and rotations are found to be strongly dependent on the temperature, the film thickness, and the distance away from the film-substrate interface. As a striking manifestation of the coupling between magnetic order and lattice structure, the Invar effect is observed below the ferromagnetic transition temperature in epitaxial layers as thin as 8 unit cells, similar to bulk ferromagnetic SrRuO$_3$.
\end{abstract}

\maketitle


\section{\label{sec:level1}Introduction}
Bulk SrRuO$_3$ (SRO) is a \textit{4d} itinerant oxide metal with a ferromagnetic Curie temperature ($T_c$) of 160 K.\cite{cao1997magnetic, mazin1997electronic, koster2012structure, klein1996transport} The system has been investigated widely as an oxide electrode in all-oxide based devices and, most recently, as a possible host of magnetic skyrmions manifesting a topological Hall effect (THE).\cite{gu2019interfacial,kimbell2020two, matsuno2016interface, ziese2020topological} Research efforts to characterize the atomic-scale structure of thin SRO films are motivated by observed thickness-dependent metal-insulator and magnetic transitions. \cite{xia2009critical,chang2009fundamental, ishigami2015thickness} A thickness-dependent metal-insulator transition (MIT) occurs in epitaxial SRO thin films fabricated by molecular beam epitaxy and pulsed laser deposition below a critical thickness of 2-4 unit cells(uc).  \cite{xia2009critical,ishigami2015thickness} The suppression of ferromagnetism in thin SRO layers is attributed to a decrease in the density of states at the Fermi level due to quantum confinement and the possible existence of an antiferromagnetic interfacial layer for ultrathin SRO films.\cite{chang2009fundamental} An important property of the SRO films grown on STO(100) is their exceptionally strong perpendicular magnetic anisotropy, preserved in films as thin as few unit cells. \cite{Schultz2009AHESROfilms, Ziese2010SROanisotropy, Wakabayashi2021SROanisotropy} This magnetic anisotropy is directly related to the intriguing observation of humplike features in Hall effect resistance loops of thin SRO films and heterostructures containing SRO layers. The humps of the Hall loops have been interpreted as a signature of the THE originating from the formation of skyrmions, as a result of a surface/interface-induced or defects-induced  Dzyaloshinskii-Moriya interaction. This is however still under debate, as  other mechanisms apart of a THE contribution can explain the occurrence of humplike features in the Hall loops as well.\cite{matsuno2016interface, Ziese2018asymmetricSRO, sohn2021stable, huang2020detection, lu2021defect, wysocki2020validity} Multiple factors are known to contribute to the anomalous Hall effect in SRO including stoichiometry, structure, temperature and layer thickness. \cite{Schultz2009AHESROfilms, Ziese2018asymmetricSRO, wysocki2020validity} Hence, a detailed understanding of the atomic-scale properties of SRO is required to understand its complex electronic and magnetic properties.

Due to the strong coupling between the lattice structure and the electronic and magnetic properties of SRO, the atomic scale structure of SRO films has been investigated theoretically and experimentally as a function of the substrate-induced strain and the coupling of oxygen octahedra across heterointerfaces, to uncover the origins of the thickness-dependent properties of the system.\cite{ziese2019unconventional, xia2009critical, lu2013control, kartik2021srotheory} For example, signatures of the THE effect have been attributed to a substrate induced local orthorhombic to tetragonal phase transition which may stabilize a chiral spin structure \cite{gu2019interfacial} and/or  magnetic inhomogeneity arising, for example, from variations in layer thickness.\cite{kimbell2020two} Thus, understanding the atomic-scale structural and interfacial interactions is crucial for decoupling intrinsic and extrinsic origins of the unique physical properties of SRO thin films. 

\begin{figure*}[ht]
\includegraphics[width=1\textwidth]{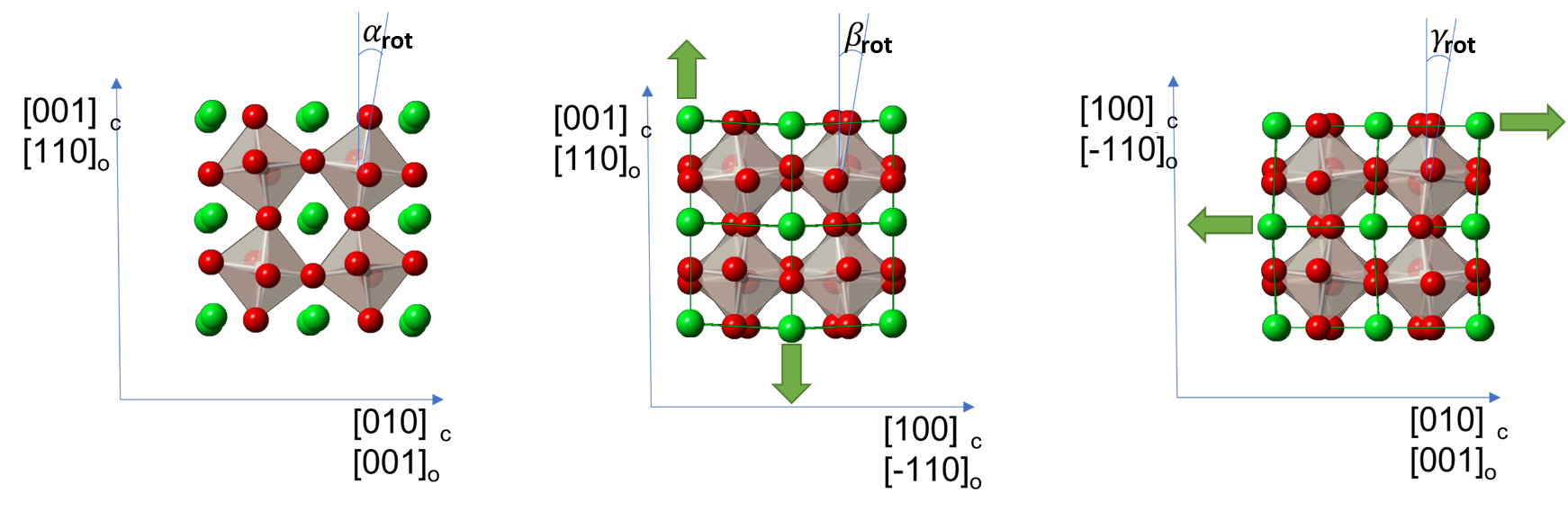} 
\caption{Schematic of structural distortions in orthorhombic SrRuO$_3$ thin films along pseudocubic (c) directions defined by the (001)-oriented SrTiO$_3$ substrate. $\alpha_{rot}$, $\beta_{rot}$ and $\gamma_{rot}$ represent the rotation angles about the [100]$_{c}$, [010]$_{c}$ and  [001]$_{c}$ axes, respectively. The orthorhombic structure is also characterized by  anti-parallel displacements of the Sr planes along  [001]$_{c}$ and  [010]$_{c}$ directions. }
\label{fig:schematic}
\end{figure*}

In this work, we report on the layer-resolved atomic scale structure of thin SRO films as a function of temperature (10 - 300 K) and film thickness (1.2-17 nm), based on high resolution synchrotron X-ray crystal truncation rod measurements (CTR) and reciprocal space maps (RSMs) and scanning transmission electron microscopy (STEM). We observe a suppression of orthorhombic/monoclinic distortions within 2-3 interfacial SRO unit cells for all film thicknesses for films grown on (001)-oriented SrTiO$_3$ substrates. The magnitude of the oxygen octahedral distortions increases on cooling between 300 K and the  ferromagnetic transition temperature (for bulk, T$_C$= 160 K) and remains constant below T$_C$ due to the Invar effect, a prominent example demonstrating the strong coupling between structure and magnetic order.\cite{kiyama1996invar}

Bulk SRO has a GdFeO$_3$-type orthorhombic crystal structure at 300 K with a$_o$=5.567 \AA{}, b$_o$= 5.530 \AA{}, c$_o$=7.845 \AA{} and $\beta=90^o$.\cite{jones1989structure} The oxygen octahedra are rotated out-of-phase along the pseudo-cubic (\textit{c}) [100]$_{c}$ and [010]$_{c}$ axes, in-phase along the [001]$_{c}$ axis(the \textit{o} and \textit{c} subscripts refer to the orthorhombic and cubic coordinates, respectively). In addition to the oxygen octahedral rotations, anti-parallel Sr cation displacements are present. Figure \ref{fig:schematic} shows the expected distortions relative to the cubic STO lattice vectors. In Glazer notation, the oxygen octahedral rotational pattern of orthorhombic SRO is represented as $a^-a^-c^+$ where the '-' superscript represents out-of-phase rotations and the '+' superscript represents in-phase rotations.\cite{glazer1972classification, woodward2005electron} At high temperatures (above 570 K), bulk SRO transitions into a tetragonal phase with in-phase rotations about the [001]$_{c}$ axis and no tilts about the [100]$_{c}$ and [010]$_{c}$ axis. The tetragonal phase is represented in Glazer notation as $a^0a^0c^+$.

Recent reports on the structure of SRO epitaxial thin films indicate that the crystal structure of SRO can be controlled by strain, the thickness of the SRO layers, the oxygen stoichiometry and the octahedral rotational pattern of the substrate.\cite{vailionis2008room, gao2016interfacial, lu2013control, aso2013atomic, chang2011thickness} SRO films compressively strained to cubic (001) SrTiO$_3$ (STO) (a=b=c= 3.905 \AA{}) are found to have a distorted orthorhombic(monoclinic) structure with the orthorhombic c-axis, [001]$_o$, parallel to the in-plane cubic [010]$_{c}$ axis of the substrate, and the orthorhombic[110]$_o$ parallel to the out-of-plane [001]$_{c}$ axis.\cite{vailionis2011misfit, gao2016interfacial} The preferential orientation of the orthorhombic axis is dictated by the step terraces.\cite{maria2000origin, rao1997growth, gan1998direct}

\section{Experimental Details}
SrRuO$_3$ samples used in our experiment were grown by
pulsed laser deposition (PLD). The samples were grown at a substrate temperature of 650-700 $^o$C and oxygen pressure of 100 mTorr, with a laser fluence of 1.5-2.0 J/cm$^2$ and a pulse repetition rate of 5 Hz. Representative atomic force microscope (AFM) images of the surface morphology of the 8, 16 and 44 unit cell (uc) SRO films are shown in supplemental Figure S1. Clear one unit cell high step terraces are resolved in the AFM images, inherent to the vicinal surface morphology of the Nb-doped (0.5 wt.$\%$ Nb) STO(100) substrates. The substrates were annealed at 925$^o$C for 1 hour in air, after they had been etched for 2.5 min in buffered HF solution.

Crystal truncation rods and half-order superstructure reflections were measured between 10 K and 300 K for 3, 8, 16 and 44 uc thick  SrRuO$_3$ films. The diffraction measurements were performed at the 33ID beamline at the Advanced Photon Source using a photon energy of 16 keV ($\lambda=0.7749 \AA{}$). The X-ray beam was focused to a spot size of 50 $\mu$m - 100 $\mu$m. The diffraction intensities were measured with a 2D Pilatus 100 K detector.\cite{schleputz2005improved} Samples were mounted in a Be-dome chamber with a base pressure of $<1 \times 10^{-5}$ Torr on a cryo-displex and the temperature was varied from 300 K to 10 K. The CTRs and half-order reflections were fit using the GenX genetic-based X-ray fitting algorithm to determine the structural properties of the films.\cite{bjorck2007genx, koohfar2017confinement} 

The structural properties of the films were also characterized with aberration corrected STEM. Samples for STEM analysis were prepared by conventional wedge polishing followed by Ar ion milling. Imaging and spectroscopy were performed on an aberration corrected FEI Titan G2 60-300 kV STEM operated at 200 kV. By simultaneously acquiring annular dark field (ADF) and integrated differential phase contrast (iDPC) images, the structure of both cationic and oxygen sublattices can be analyzed. The revolving STEM (RevSTEM) imaging technique\cite{sang2014revolving} was used to improve the signal-to-noise ratio to allow for the resolution of the shape of the atom columns revealing structural features beyond atom column positions. The film structure was determined by fitting each atom column to a two-dimensional Gaussian function allowing for a determination of the column amplitude (point image intensity), \textit{x} (in-plane) - and \textit{y} (out-of-plane)-positions, as well as the \textit{x} and \textit{y}-components of peak widths, $\sigma_x$ and $\sigma_y$. The local distortions and tilting of oxygen octahedra are characterized by the oxygen column positions and ellipticity, defined as $ E =\frac{\sigma_x}{\sigma_y}$. With the ellipticity, we characterize the out-of-phase octahedral tilting along a column.

\section{Results and Discussion}

\begin{figure}[ht]
\includegraphics[width=0.5\textwidth]{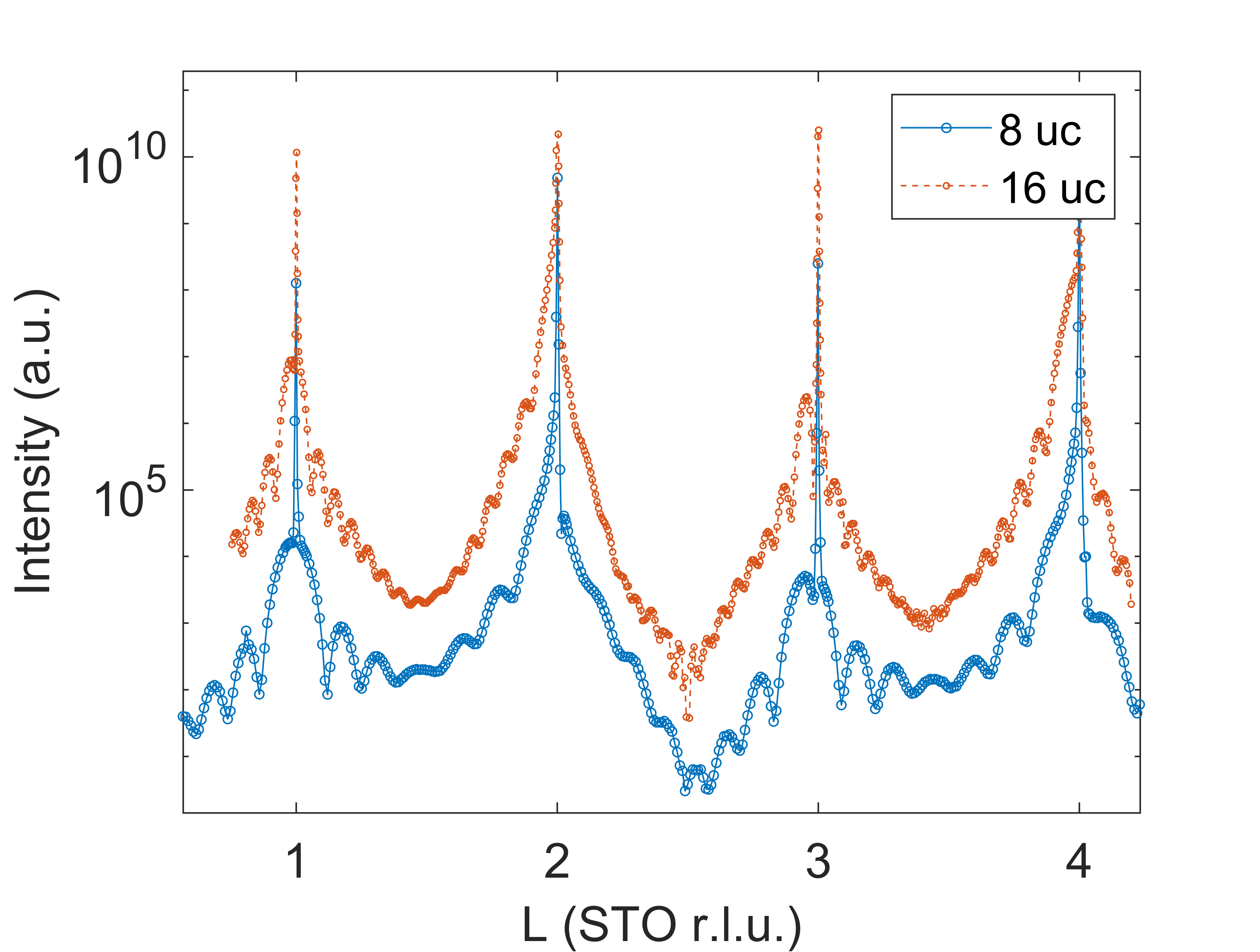} 
\caption{Specular 00L scans for 8 and 16 unit cell thick SRO films on STO. The plots are offset vertically for clarity.}
\label{fig:specular}
\end{figure}

\begin{figure*}[ht]
\includegraphics[width=1\textwidth]{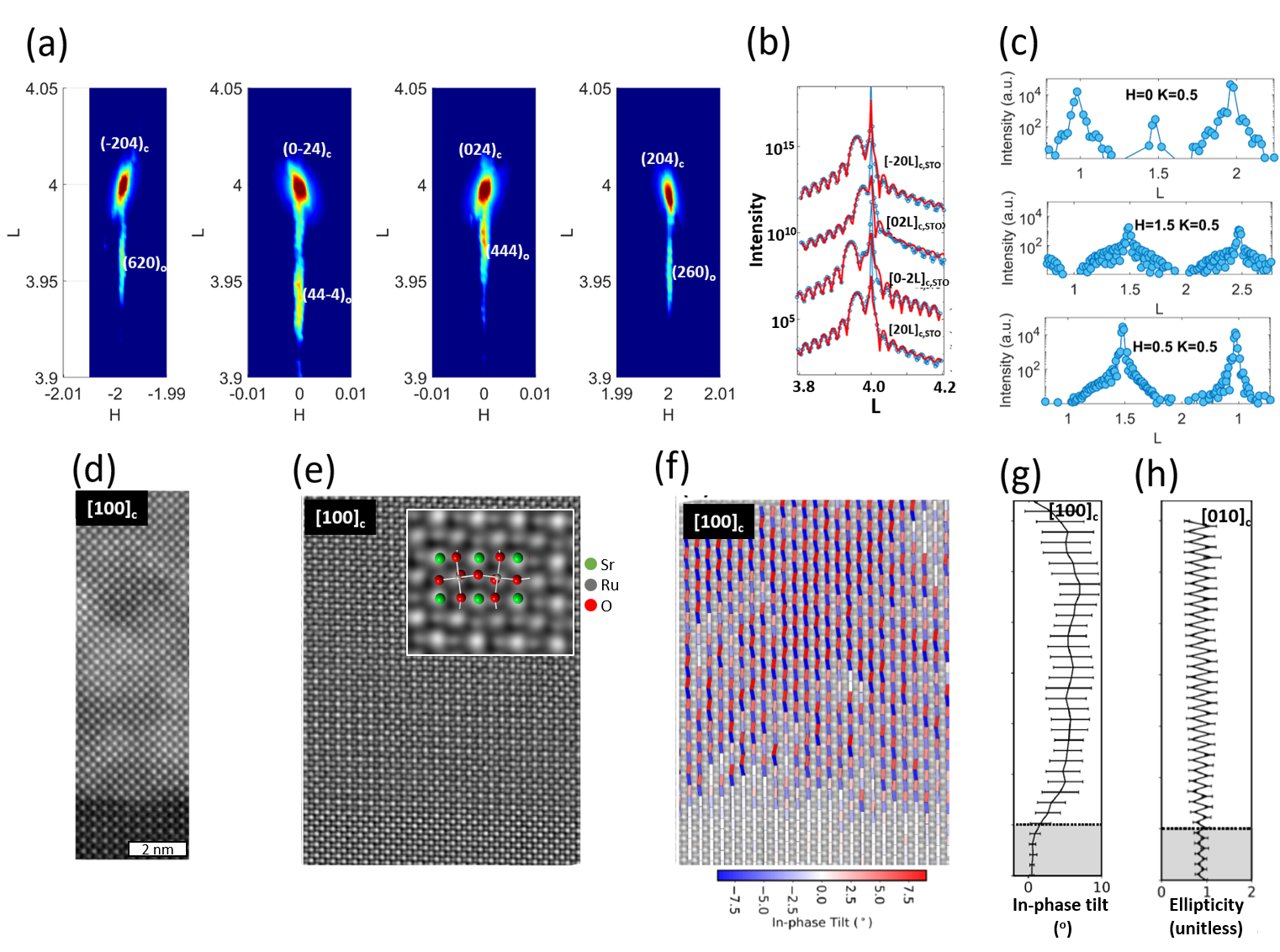} 
\caption{Reciprocal space map around the STO [204]$_{c}$ Bragg peak showing the in-equivalent orthorhombic SRO Bragg peaks for a 44 uc SRO film. (1 STO reciprocal lattice unit = 1/0.3905 nm$^{-1}$) (b) Measured (blue circle) line profiles along the STO L direction in (a) and fits (solid red lines). (c) Measured half-order rods (d) Annular dark field images of the 44 uc SRO film along the   [100]$_c$ projections with  (e) corresponding iDPC images. The inset shows a magnification of the iDPC image with the XRD structure overlayed. (f) Measured in-phase tilting angle map along the [100] zone axis with respective layer-resolved (g) in-phase tilt angles and h) ellipticity  profiles.}
\label{fig:Structure_44uc}
\end{figure*}

Figure \ref{fig:specular} shows the measured specular (00L) CTRs for the 8 and 16 uc samples. The presence of finite thickness oscillations is indicative of flat surfaces and a chemically abrupt SRO/STO interface. The out-of-plane lattice parameters determined from fits to the (00L) CTRs for the 8, 16 and 44 uc SRO films are 3.957,  3.955  and 3.948 \AA{}, respectively. The increased lattice spacing compared to the bulk pseudocubic value for SRO (a$_{c, bulk}=3.93 \AA{}$) is due to the biaxial compressive in-plane strain imposed by the STO substrate. The films are coherently strained to the STO substrate with an in-plane pseudocubic lattice constant of 3.905 \AA{} as evidenced by RSMs measured by X-ray diffraction (Figure \ref{fig:Structure_44uc}(a)).

The reduced symmetry of the thickest (44 uc) SRO film relative to the cubic \textit{Pm$\bar{3}$m} STO substrate is observed in  RSMs measured around the  STO (204)$_c$,(024)$_c$,(-204)$c$ and (0-24)$_c$ Bragg peaks. The measured RSMs are compared in Figure \ref{fig:Structure_44uc}(a) where 1 STO reciprocal lattice unit (r.l.u.) = 1/0.3905 nm$^{-1}$. The different \textit{L} values for the SRO (6,2,0)$_o$ and (2,6,0)$_o$ Bragg peaks with fixed values of Q$_{in-plane}$ is due to the monoclinic distortion of the SRO lattice.\cite{vailionis2011misfit} The angle $\beta_{mon}$, between the $a_o^{SRO}$ and $b_o^{SRO}$ axes is determined from the relation $\beta_{mon}=90-arctan(\frac{\Delta Q_z}{Q_{inplane}})$\cite{gao2016interfacial}. From the RSMs we determine the monoclinic SRO lattice parameters to be $\beta_{mon}=89.5^o$, $a_o^{SRO}=5.72 \AA{}$, $b_o^{SRO}=5.52 \AA{}$ and $c_o^{SRO}=7.81 \AA{}$. The lattice parameters are also confirmed from fits to the in-equivalent (04L)$_c$ CTRs and half order peaks in Figure \ref{fig:Structure_44uc}(b) and \ref{fig:Structure_44uc}(c).

\begin{figure*}[ht]
\includegraphics[width=1\textwidth]{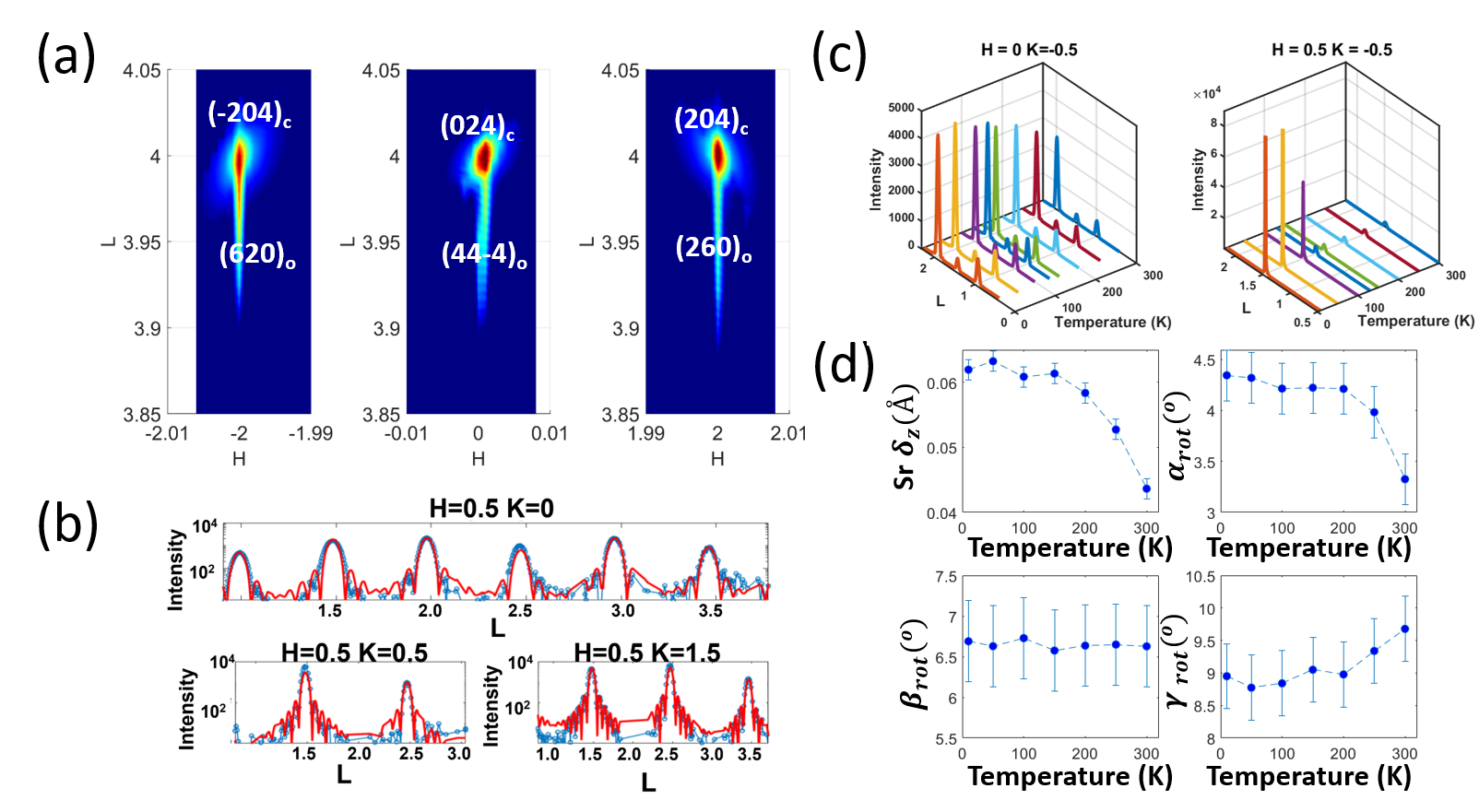} 
\caption{(a) Reciprocal space maps for the 16 uc SRO/STO sample.(b) Measured (blue circle) half order rod and fits (solid red lines).(c) Intensity of half order peaks as the function of temperature.(d) Sr displacement and octahedral rotation angles as functions of temperature.}
\label{fig:Strucutre16}
\end{figure*}

The RSMs for the 16 uc film around the STO $\{204\}_c$ Bragg peaks are shown in Figure \ref{fig:Strucutre16}(a). The film Bragg peaks for the 16 uc sample are broader along the \textit{L} direction than the 44 uc sample due to  the effect of the finite thickness broadening and the increased fraction of the multiple rotational domains. Thus, care must be taken in relying solely on RSMs in verifying the orthorhombicity of the lattice.

The octahedral rotations about the out-of-plane [001]$_c$ axis and tilts about the in-plane [100]$_c$ and [010]$_c$ axes can be qualitatively predicted from the prescence/absence of half-order reflections. A c$^+$ in-phase rotation along the [001]$_c$ axis results in reflections of type $\frac{1}{2}$(\textit{odd,odd,even})$_{c}$.\cite{woodward2005electron, May2010LaNiO3} The absence of peaks at integer L along the (3/2, 1/2,  L)$_c$ rod for the 44 uc film in Figure \ref{fig:Structure_44uc}(c) indicates that the axis with in-phase rotations does not lie along the out-of-plane [001]$_c$ axis.

Bragg peaks are expected for an $a^+b^-c^-$ structure 
for reflections of type $\frac{1}{2}$(\textit{even,odd,even})$_{c}$ and $\frac{1}{2}$(\textit{odd,odd,odd})$_{c}$. The (0 1/2 \textit{even})$_c$ peaks observed in Figure \ref{fig:Structure_44uc}(c) indicate an a$^+$ tilt. Based on the observed reflections for the 44 uc film, the tilt system of the SRO on STO is determined to be $a^+b^-c^-$. However, the presence of reflections of type $\frac{1}{2}$(\textit{even, odd, odd})$_{c}$ signifies the presence of an $a^-b^+c^-$ domain with the in-phase tilt along the [010]$_c$ direction.

Due to the four-fold symmetry of the cubic STO substrate, 4 rotational variants of the SRO unit cell are  expected with the [001]$_o$ axis aligned along either the STO [100]$_c$,[-100]$_c$,[010]$_c$ or[0-10]$_c$ axis.\cite{gao2016interfacial, vailionis2011misfit, siwakoti2021abrupt} The half order peaks associated with the $a^-b^+c^-$ domain are relatively weak for the 44 uc sample and the fraction of the film with the $a^-b^+c^-$ orientation is less than 5\% of the total film volume. However, for the 16 uc sample, the ratio of the $a^+b^-c^-$ domain fraction to the $a^-b^+c^-$ domain is 3:1 suggesting that the film transitions into a single domain structure as the thickness increases or slight differences exist in the miscut angles and direction of the STO substrates. \cite{maria2000origin, wang2020magnetic, kar2021high} Indeed topographic atomic force microscopy investigations (supplemental Fig. S1) showed that the STO(100) substrate used for the 16 uc sample has tilted terraces with respect to its edges, while the one used for the 44 uc has terraces running almost parallel to the substrate edges. It was shown that more tilted terraces promote the formation of crystallographic domains with different in-plane orientations of the long orthorhombic \textit{c} axis.\cite{wang2020magnetic}

\begin{table}[]
    \centering
    \begin{tabular}{|c|c|c|c|c|}
     \hline
         Parameters                   &8 uc                   &	16 uc    	     & 44uc  \\
         \hline 
          c (\AA{})                           &3.957              &	3.955           & 3.948 \\ \hline

          $\alpha_{rot}$ (in-phase), 300 K           &0.9$^o$                  & 3.4$^o$                & 5.31$^o$     \\
          $\beta_{rot}$ (out-of-phase), 300 K        &3.3$^o$                  & 6.4 $^o$              &  5.72$^o$     \\
          $\gamma_{rot}$ (out-of-phase), 300 K       &9.9$^o$                  & 9.7$^o$               &  5.21$^o$     \\
          \hline
          $\alpha_{rot}$ (in-phase),130 K          &1.7$^o$                 & 4.2$^o$               & -     \\
          $\beta_{rot}$ (out-of-phase),130 K       &2.6$^o$                 & 6.6$^o$               & -    \\
          $\gamma_{rot}$ (out-of-phase) 130 K      &9.6$^o$                 & 9.1$^o$               & -    \\
          \hline
          $Ru-O-Ru $           &159$^o$                       & 162$^o$               & 164$^o$     \\
          $Sr \delta_z A$ (300 K)          &0.009                       & 0.040               & 0.062    \\
          $Sr \delta_z A$ (130 K)          &0.035                       & 0.060               & -    \\

            \hline
       
    \end{tabular}
    \caption{Comparison of structural parameters of non-interfacial and non-surface layers of SRO films as a function of thickness and temperature. }
    \label{tab:fitresults}
\end{table}

The local structure of the 44 uc film is investigated by STEM measurements shown in Figure \ref{fig:Structure_44uc}(d)-(h). The ADF image in Figure \ref{fig:Structure_44uc}(d) indicates a chemically abrupt interface between the SRO film and the STO substrate. The oxygen sublattice is imaged using the iDPC technique. Along the [100]$_c$ projection where the oxygen octahedra rotate in-phase, the tilt angles are determined directly from the atomic positions of the oxygen atoms in the iDPC image in Figure \ref{fig:Structure_44uc}(e). Figure \ref{fig:Structure_44uc}(f) shows a map of the unit-cell resolved oxygen octahedral tilts about the [100]$_c$ axis. A depth profile of the magnitude of the layer-averaged in-phase rotation angle is shown in Figure \ref{fig:Structure_44uc}(g). There are no in-phase rotations in the STO substrate as expected for bulk STO. The rotation angle increases gradually from 0 to 5$^\circ$ within the three interfacial SRO layers and remains uniform till the surface layer where a reduction occurs. The suppressed in-phase tilt at the film-substrate interface is consistent with fits to the X-ray diffraction data discussed below and previous reports.\cite{siwakoti2021abrupt}

A projection along the [010]$_c$ direction where the octahedra rotate out-of-phase results in an asymmetric smearing of the oxygen atomic columns. Thus, the projected out-of-phase octahedral tilting is determined by the ellipticity $E=\sigma_x/\sigma_y$ of the atomic columns, where $\sigma_x$ and $\sigma_y$ represent the horizontal and vertical peak widths measured from fitting the iDPC images. The layer-averaged ellipticity profile is shown in Figure \ref{fig:Structure_44uc}(h) along [010]$_c$ projection. 
In the \textit{AO} planes, $E<1$ since the rotations lead to O displacements in the vertical direction. Conversely, $E>1$ in the \textit{$BO_2$} planes due to O displacements in the vertical direction. $E=1$ corresponds to no rotations. The alternating values of $E$ about 1 in Figure \ref{fig:Structure_44uc}(h) corresponds to successive \textit{SrO} and \textit{$BO_2$} (\textit{B}=Ti, Ru) planes. Some ellipticity is measured in the STO substrate, likely due to low order aberrations and/or rotations in the interfacial STO layers induced by the SRO adlayer.

The iDPC images provide a direct quantitative measure of the magnitude of the in-phase rotations about the [100]$_c$ axis. To quantitatively determine the layer-resolved rotations about the [010]$_c$ and [001]$_c$ axes and the  orthorhombic Sr displacements, the CTR's and half-order rods measured by synchrotron X-ray diffraction are analyzed using the GenX genetic fitting algorithm.\cite{bjorck2007genx} To account for differences in the structure of the interfacial 3 unit cells, separate fit (rotation angles, Sr displacements) parameters are assigned to the interface layers and the non-interfacial layers. 

Table \ref{tab:fitresults} summarizes the structural parameters determined for the SRO films as a function of thickness at 300 K and 130 K. The measured octahedral rotation angles for the  non-interfacial layers of the 44 uc film about the orthogonal cubic axes are $\alpha_{rot}=5.3^o$, $\beta_{rot} = 5.7^o$, $\gamma_{rot} = 5.2^o$ in good agreement with bulk values of 6.19$^o$, 5.97$^o$ and 5.97$^o$ respectively. A suppression of tilts and Sr displacement is found at the interface due to the structural coupling to the cubic STO substrate which possesses no octahedral rotations at 300 K. This is consistent  with the iDPC results in Figure \ref{fig:Structure_44uc}(g). No evidence for off-center oxygen displacements in the STO and SRO were found as has been recently reported in 4 uc SRO films. \cite{sohn2021stable}

For the 16 uc SRO film at 300 K, the in-phase rotation angle is suppressed to 3.4$^o$ while the rotation about c-axis increases (relative to bulk) to 9.7$^o$. The distorted orthorhombic structure at 300 K is in contrast to the tetragonal a$^0$b$^0$c$^-$ structure  reported for SRO films with thicknesses below 17 uc.\cite{chang2011thickness} Since oxygen vacancies can stabilize the tetragonal structure, the discrepancy is most likely related to the oxygen stoichiometry.

The evolution of the structure of the 16 uc film with temperature between 10 K and 300 K is determined from fits to the temperature-dependent half order rods. Representative measured  half order rods and fits at 300 K for the 16 uc film are shown in Figure \ref{fig:Strucutre16}(b). The intensity of the (0 -0.5 L)$_c$ and (0.5 -0.5 1.5)$_c$ peaks measured as a function of temperature from 300 K to 10 K for the 16 uc film are shown in Figure \ref{fig:Strucutre16}(c). The intensities of the half order peaks associated with the in-phase octahedral tilts and Sr displacements increase as the temperature decreases to the FM-PM transition at $\approx$150 K. Below 150 K, the intensity of the (0 -0.5 L)$_c$ peaks  remains constant indicating a freezing of the octahedral distortions in the ferromagnetic phase. \cite{el2011modeling}  A sharp increase in the intensity of the (0.5 -0.5 L)$_c$ peaks is observed below 105 K where the antiferrodistortive STO phase transition occurs. The STO phase transition involves rotation of the TiO$_6$ octahedra in the STO substrate leading to a doubling of the STO unit cell and the emergence of the STO substrate $\frac{1}{2}$(\textit{odd, odd,odd})$_c$ peaks. The temperature-dependent structural parameters for the 16 uc SRO film are summarized in Figure \ref{fig:Strucutre16}(d). The Sr displacements in the SRO layers increase from 0.04 \AA{} at 300 K to 0.06 \AA{} at 150 K and the in-phase rotation angle increases from 3.3$^o$ at 300 K to 4.2$^o$ at 150 K.

In contrast to the thicker films, the intensity of of the half-order reflections for the 8 uc film associated with the in-phase tilts and Sr displacements are strongly suppressed at 300 K. The STEM analysis and the XRD results are described in Figure S3 and S4 of the supplemental materials. The rotation angles at 300 K away from the film-substrate interface are $\alpha_{rot}=0.9 ^o$, $\beta_{rot}=3.3 ^o$ and $\gamma_{rot}=9.9 ^o$. The suppression of the in-plane tilts and the enhancement of the rotation around the c-axis relative to bulk leads to an in-plane Ru-O-Ru bond angle of 159$^o$ which is slightly less than the value of 162$^\circ$ for bulk orthorhombic SRO.  Fits to the half-order reflections at 130 K in the ferromagnetic phase show a slight increase in the in-phase tilt angle to 1.7 $^o$ but still less than the expected value for bulk SRO. 

\begin{figure}[ht]
\includegraphics[width=0.5\textwidth]{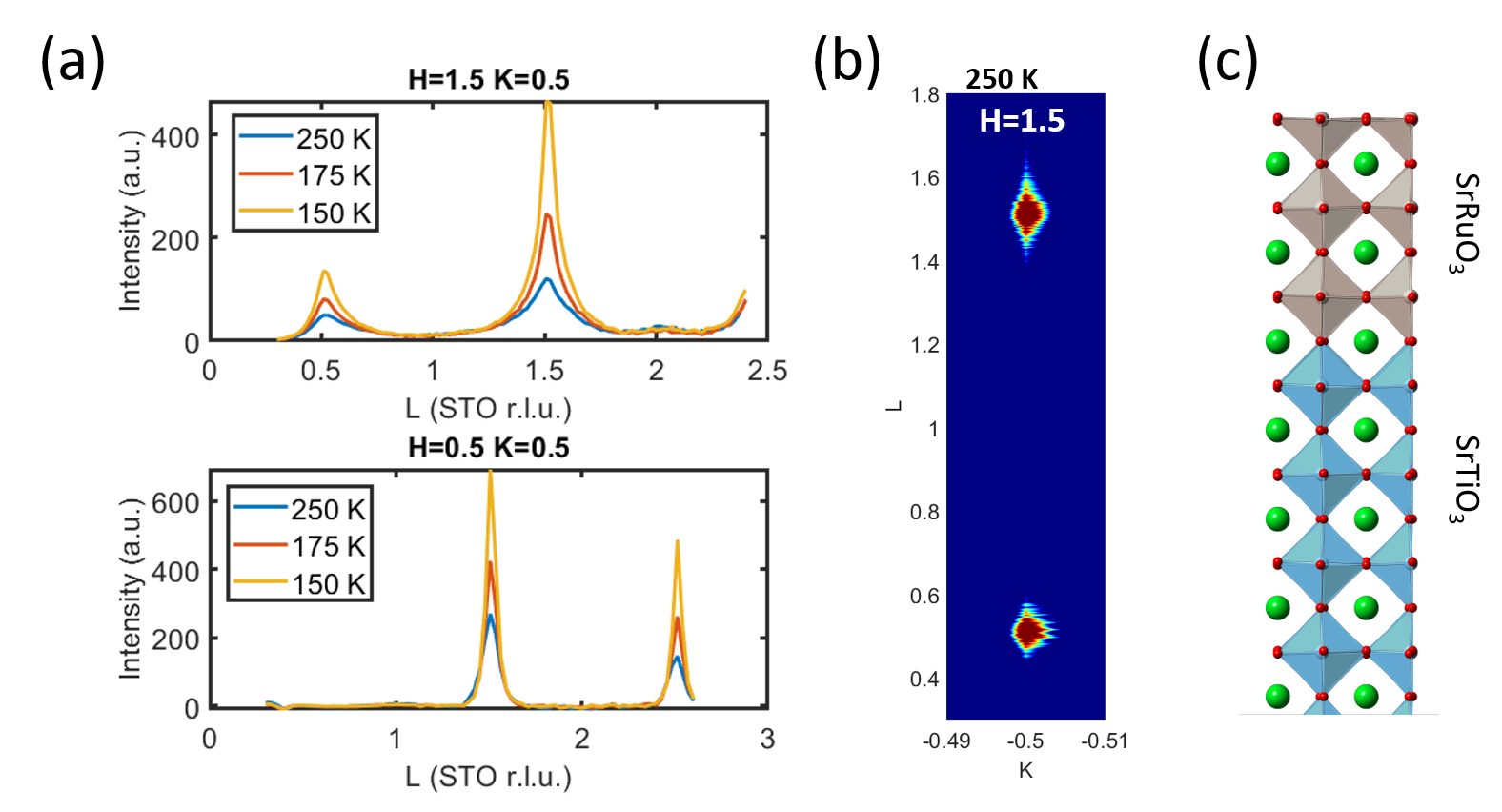} 
\caption{{(a) Measured half-order rods as a function of temperature for a 3 uc SRO film. (b) Reciprocal space map along the H=1.5 K=0.5 L rod and (c) Model structure of rotations in SRO film and interfacial STO layers.}}

\label{fig:S3uc}
\end{figure}

Octahedral distortions are also observed for the thinnest 3 uc SRO sample. Half-order Bragg peaks are observed for $\frac{1}{2}$(\textit{odd}, \textit{odd}, \textit{odd}) reflections for the 3 uc SRO sample while peaks associated with in-phase rotations and Sr displacements are absent between 300 K and 10 K. A $a^0a^0c^-$ tetragonal film structure can be ruled out since (1/2 1/2 $odd/2$) diffraction peaks are present as shown in Figure \ref{fig:S3uc}(a) which indicate additional tilts about the [001]$_c$ and [010]$_c$ axes. The peak intensities increase as the temperature decreases from 250 K to 150 K indicating an increase in the magnitude of the octahedral tilts and rotations with decreasing temperature. The tilts and rotations observed in the thinnest sample are consistent with low energy electron diffraction results on single unit cell thick SRO films reported by Siwakoti \textit{et al}.\cite{siwakoti2021abrupt}  The peak widths along the L direction in the RSM image in Figure \ref{fig:S3uc} is narrower than width expected for a 3 uc thick film suggesting that the octahedral rotations extend 2-3 layers into the STO substrate as it was observed for CaRuO$_3$ and (La,Sr)MnO$_3$ films on (001)-SrTiO$_3$ by STEM imaging.\cite{siwakoti2021abrupt, koohfar2017confinement} The induced rotations in the STO are found to be $<$ 1$^o$ in magnitude and out-of-phase around the [010] and [001] axes and thus, will be difficult to observe directly in ABF images. However, the half-order reflections obtained by synchrotron X-ray diffraction measurements are sensitive to these distortions.

The thickness and temperature-dependent structural results are summarized as follows: 1) a suppression of orthorhombic distortions occurs within the 2-3 SRO layers at the SRO/STO interface consistent with theoretical predictions\cite{he2010control}; 2) reduced A-site displacements are correlated with a decrease in the magnitude of in-phase rotations and an increase in the oxygen octahedral rotations about the c-axis as the film thickness is reduced from 44 uc to 8 uc; 3) freezing of rotations and A-site displacements occurs below the paramagnetic-ferromagnetic transition temperature.

The suppression of octahedral rotations about the orthogonal in-plane axis for the SRO layers close to the SRO/STO interface is expected due to structural coupling with the cubic STO substrate and these results are consistent with previous reports on the SRO/STO interface structure. We find that the  orthorhombic distortions are present, but weak in the 8 uc film, with the magnitude of the distortions increasing with decreasing temperature.

The freezing out of distortions in the 8, 16 and 44 uc films below T$_C$  arises due to the coupling of the magnetic moment ordering to the lattice structure. This effect which is also known as the Invar effect is characterized by a freezing out of the unit cell volume and octahedral rotations below the ferromagnetic transition temperature, leading to anomalously low coefficient of thermal expansion: This effect was observed in bulk SRO and our results confirm the existence of the Invar effect in ultrathin strained SRO films.\cite{dabrowski2006magnetic, kiyama1996invar, bushmeleva2006evidence}  

We find that, while the 8 uc sample is close to an orthorhombic-tetragonal transition at 300 K, the orthorhombic distortions become stronger as the temperature decreases and the low temperature structure, where humplike anomalies of the Hall resistance loops were observed and taken for a THE fingerprint \cite{gu2019interfacial}, is orthorhombic.  Based on  structural investigations done at 300 K, Gu \textit{et. al.} proposed that in their 8 uc SRO films,  the interfacial RuO$_6$ octahedral tilting induced by a local orthorhombic-to-tetragonal structural phase transition across the SRO/STO interface resulted in breaking the inversion symmetry. This was further used to account for an interfacial DMI and to explain why a THE contribution may occur in the Hall resistance loops \cite{gu2019interfacial}. Our findings are therefore very important, because we have information about the structure at the temperature where the physical properties are measured. The temperature and thickness dependent structures of SRO layers is highly relevant for understanding the magnetocrystalline anisotropy, the anomalous Hall effect and the temperature dependence and sign of the anomalous Hall constant for tetragonal and orthorhombic SRO phases. \cite{Kan2013tetragonalSRO, ziese2019unconventional, Bern2013AHE, kartik2021srotheory}

In addition to the structural distortions observed in this work, local variations in film thickness arising from the step-flow-growth mechanism can lead non-uniform coercivity and humplike features in magnetoresistance measurements which mimick the THE. \cite{malsch2020correlating, kimbell2020two} Magnetic force measurements indicate lateral variations in the coercivity of nominally 4 uc films capped with SrIrO$_3$/SrZrO$_3$ layers which may be correlated with local variations in the film thickness. For the thicker films (8-44 uc), the surface occupation of the top 2-3 layers is slightly less than unity, indicating that the local film thickness averaged over the X-ray probe area (100 $\mu$m) fluctuates within this range. These variations become more critical as the film thickness is reduced below 8 unit cells. 

While this analysis assumes no structural or chemical changes occur upon the surface exposure to ambient conditions, in some oxides such changes can be quite significant.\cite{Caspi2022SrVO3} SRO films exposed to ambient atmosphere and heated in vacuum may decompose.\cite{SHIN2005SROsurface-stability} However, the strong signal originating both from the half order reflections and the RSMs (Figure \ref{fig:S3uc}) indicate that even at 3 uc, a significant portion of the film retains its high quality crystalline structure. Overoxidation and other surface reactions could sometimes result in amorphous phases, which would not be observable in diffraction experiments. Nonetheless, the strong diffraction intensity (originating from inherently low-intensity reflections) indicates that this possibility is not playing a significant role in the current case.

\section{Conclusion}
In summary, the atomic-scale structure of SRO films was investigated as a function of temperature and film thickness. The ultrathin films (3 and 8 uc thick) show the most pronounced structural changes with respect to the bulk structure. An increase of the magnitude of the oxygen octahedral rotations and Sr displacements is observed as the temperature is reduced from 300 K to the ferromagnetic transition, below 130 K. The freezing out of the octahedral distortions observed below 130 K in the films as thin as 8 unit cells is associated with the Invar effect, which is known to occur in bulk ferromagnetic SRO, as a consequence of the coupling between the crystal structure and ferromagnetic ordering. The thickness-dependent structure of the SRO films may be related to kinetic and thermodynamic effects during nucleation of the orthorhombic structure of SRO thin film on the cubic STO substrate. Further structural investigations shall enable an \textit{in depth} understanding of the unique properties of SRO thin films, with the particular motivation of shedding light on the intriguing magnetotransport properties reported for ultra-thin films.

\section*{Acknowledgments}
The authors acknowledge financial support by the US National Science Foundation under Grant No. NSF DMR-1751455. This work was performed in part at the Analytical Instrumentation Facility (AIF) at North Carolina State University, which is supported by the State of North Carolina and the National Science Foundation (award number ECCS-2025064). The AIF is a member of the North Carolina Research Triangle Nanotechnology Network (RTNN), a site in the National Nanotechnology Coordinated Infrastructure (NNCI). This material is based upon work supported by the National Science Foundation under Grant No. DGE-1633587. We thank Ren\'{e} Borowski and Silvia de Waal for etching the STO substrates and Regina Dittmann and Felix Gunkel for access to the PLD system at FZ Jülich. I.L.-V. acknowledges the financial support from the German Research Foundation (DFG) for project no. 403504808 within SPP2137 and for project no. 277146847 within SFB1238 (project A01). L. K. thanks German Israeli Foundation for financial support (GIF Grant no. I-1510-303.10/2019). Use of the Advanced Photon Source was supported by the U.S. Department of Energy, Office of Science, Office of Basic Energy Sciences, under Contract No. DE-AC02-06CH11357.


\begin{thebibliography}{51}%
\makeatletter
\providecommand \@ifxundefined [1]{%
 \@ifx{#1\undefined}
}%
\providecommand \@ifnum [1]{%
 \ifnum #1\expandafter \@firstoftwo
 \else \expandafter \@secondoftwo
 \fi
}%
\providecommand \@ifx [1]{%
 \ifx #1\expandafter \@firstoftwo
 \else \expandafter \@secondoftwo
 \fi
}%
\providecommand \natexlab [1]{#1}%
\providecommand \enquote  [1]{``#1''}%
\providecommand \bibnamefont  [1]{#1}%
\providecommand \bibfnamefont [1]{#1}%
\providecommand \citenamefont [1]{#1}%
\providecommand \href@noop [0]{\@secondoftwo}%
\providecommand \href [0]{\begingroup \@sanitize@url \@href}%
\providecommand \@href[1]{\@@startlink{#1}\@@href}%
\providecommand \@@href[1]{\endgroup#1\@@endlink}%
\providecommand \@sanitize@url [0]{\catcode `\\12\catcode `\$12\catcode
  `\&12\catcode `\#12\catcode `\^12\catcode `\_12\catcode `\%12\relax}%
\providecommand \@@startlink[1]{}%
\providecommand \@@endlink[0]{}%
\providecommand \url  [0]{\begingroup\@sanitize@url \@url }%
\providecommand \@url [1]{\endgroup\@href {#1}{\urlprefix }}%
\providecommand \urlprefix  [0]{URL }%
\providecommand \Eprint [0]{\href }%
\providecommand \doibase [0]{http://dx.doi.org/}%
\providecommand \selectlanguage [0]{\@gobble}%
\providecommand \bibinfo  [0]{\@secondoftwo}%
\providecommand \bibfield  [0]{\@secondoftwo}%
\providecommand \translation [1]{[#1]}%
\providecommand \BibitemOpen [0]{}%
\providecommand \bibitemStop [0]{}%
\providecommand \bibitemNoStop [0]{.\EOS\space}%
\providecommand \EOS [0]{\spacefactor3000\relax}%
\providecommand \BibitemShut  [1]{\csname bibitem#1\endcsname}%
\let\auto@bib@innerbib\@empty
\bibitem [{\citenamefont {Cao}\ \emph {et~al.}(1997)\citenamefont {Cao},
  \citenamefont {McCall}, \citenamefont {Shepard}, \citenamefont {Crow},\ and\
  \citenamefont {Guertin}}]{cao1997magnetic}%
  \BibitemOpen
  \bibfield  {author} {\bibinfo {author} {\bibfnamefont {G.}~\bibnamefont
  {Cao}}, \bibinfo {author} {\bibfnamefont {S.}~\bibnamefont {McCall}},
  \bibinfo {author} {\bibfnamefont {M.}~\bibnamefont {Shepard}}, \bibinfo
  {author} {\bibfnamefont {J.}~\bibnamefont {Crow}}, \ and\ \bibinfo {author}
  {\bibfnamefont {R.}~\bibnamefont {Guertin}},\ }\href@noop {} {\bibfield
  {journal} {\bibinfo  {journal} {Physical Review B}\ }\textbf {\bibinfo
  {volume} {56}},\ \bibinfo {pages} {R2916} (\bibinfo {year}
  {1997})}\BibitemShut {NoStop}%
\bibitem [{\citenamefont {Mazin}\ and\ \citenamefont
  {Singh}(1997)}]{mazin1997electronic}%
  \BibitemOpen
  \bibfield  {author} {\bibinfo {author} {\bibfnamefont {I.}~\bibnamefont
  {Mazin}}\ and\ \bibinfo {author} {\bibfnamefont {D.~J.}\ \bibnamefont
  {Singh}},\ }\href@noop {} {\bibfield  {journal} {\bibinfo  {journal}
  {Physical Review B}\ }\textbf {\bibinfo {volume} {56}},\ \bibinfo {pages}
  {2556} (\bibinfo {year} {1997})}\BibitemShut {NoStop}%
\bibitem [{\citenamefont {Koster}\ \emph {et~al.}(2012)\citenamefont {Koster},
  \citenamefont {Klein}, \citenamefont {Siemons}, \citenamefont {Rijnders},
  \citenamefont {Dodge}, \citenamefont {Eom}, \citenamefont {Blank},\ and\
  \citenamefont {Beasley}}]{koster2012structure}%
  \BibitemOpen
  \bibfield  {author} {\bibinfo {author} {\bibfnamefont {G.}~\bibnamefont
  {Koster}}, \bibinfo {author} {\bibfnamefont {L.}~\bibnamefont {Klein}},
  \bibinfo {author} {\bibfnamefont {W.}~\bibnamefont {Siemons}}, \bibinfo
  {author} {\bibfnamefont {G.}~\bibnamefont {Rijnders}}, \bibinfo {author}
  {\bibfnamefont {J.~S.}\ \bibnamefont {Dodge}}, \bibinfo {author}
  {\bibfnamefont {C.-B.}\ \bibnamefont {Eom}}, \bibinfo {author} {\bibfnamefont
  {D.~H.}\ \bibnamefont {Blank}}, \ and\ \bibinfo {author} {\bibfnamefont
  {M.~R.}\ \bibnamefont {Beasley}},\ }\href@noop {} {\bibfield  {journal}
  {\bibinfo  {journal} {Reviews of Modern Physics}\ }\textbf {\bibinfo {volume}
  {84}},\ \bibinfo {pages} {253} (\bibinfo {year} {2012})}\BibitemShut
  {NoStop}%
\bibitem [{\citenamefont {Klein}\ \emph {et~al.}(1996)\citenamefont {Klein},
  \citenamefont {Dodge}, \citenamefont {Ahn}, \citenamefont {Reiner},
  \citenamefont {Mieville}, \citenamefont {Geballe}, \citenamefont {Beasley},\
  and\ \citenamefont {Kapitulnik}}]{klein1996transport}%
  \BibitemOpen
  \bibfield  {author} {\bibinfo {author} {\bibfnamefont {L.}~\bibnamefont
  {Klein}}, \bibinfo {author} {\bibfnamefont {J.}~\bibnamefont {Dodge}},
  \bibinfo {author} {\bibfnamefont {C.}~\bibnamefont {Ahn}}, \bibinfo {author}
  {\bibfnamefont {J.}~\bibnamefont {Reiner}}, \bibinfo {author} {\bibfnamefont
  {L.}~\bibnamefont {Mieville}}, \bibinfo {author} {\bibfnamefont
  {T.}~\bibnamefont {Geballe}}, \bibinfo {author} {\bibfnamefont
  {M.}~\bibnamefont {Beasley}}, \ and\ \bibinfo {author} {\bibfnamefont
  {A.}~\bibnamefont {Kapitulnik}},\ }\href@noop {} {\bibfield  {journal}
  {\bibinfo  {journal} {Journal of Physics: Condensed Matter}\ }\textbf
  {\bibinfo {volume} {8}},\ \bibinfo {pages} {10111} (\bibinfo {year}
  {1996})}\BibitemShut {NoStop}%
\bibitem [{\citenamefont {Gu}\ \emph {et~al.}(2019)\citenamefont {Gu},
  \citenamefont {Wei}, \citenamefont {Xu}, \citenamefont {Zhang}, \citenamefont
  {Wang}, \citenamefont {Li}, \citenamefont {Saleem}, \citenamefont {Chang},
  \citenamefont {Sun}, \citenamefont {Song} \emph
  {et~al.}}]{gu2019interfacial}%
  \BibitemOpen
  \bibfield  {author} {\bibinfo {author} {\bibfnamefont {Y.}~\bibnamefont
  {Gu}}, \bibinfo {author} {\bibfnamefont {Y.-W.}\ \bibnamefont {Wei}},
  \bibinfo {author} {\bibfnamefont {K.}~\bibnamefont {Xu}}, \bibinfo {author}
  {\bibfnamefont {H.}~\bibnamefont {Zhang}}, \bibinfo {author} {\bibfnamefont
  {F.}~\bibnamefont {Wang}}, \bibinfo {author} {\bibfnamefont {F.}~\bibnamefont
  {Li}}, \bibinfo {author} {\bibfnamefont {M.~S.}\ \bibnamefont {Saleem}},
  \bibinfo {author} {\bibfnamefont {C.-Z.}\ \bibnamefont {Chang}}, \bibinfo
  {author} {\bibfnamefont {J.}~\bibnamefont {Sun}}, \bibinfo {author}
  {\bibfnamefont {C.}~\bibnamefont {Song}},  \emph {et~al.},\ }\href@noop {}
  {\bibfield  {journal} {\bibinfo  {journal} {Journal of Physics D: Applied
  Physics}\ }\textbf {\bibinfo {volume} {52}},\ \bibinfo {pages} {404001}
  (\bibinfo {year} {2019})}\BibitemShut {NoStop}%
\bibitem [{\citenamefont {Kimbell}\ \emph {et~al.}(2020)\citenamefont
  {Kimbell}, \citenamefont {Sass}, \citenamefont {Woltjes}, \citenamefont {Ko},
  \citenamefont {Noh}, \citenamefont {Wu},\ and\ \citenamefont
  {Robinson}}]{kimbell2020two}%
  \BibitemOpen
  \bibfield  {author} {\bibinfo {author} {\bibfnamefont {G.}~\bibnamefont
  {Kimbell}}, \bibinfo {author} {\bibfnamefont {P.~M.}\ \bibnamefont {Sass}},
  \bibinfo {author} {\bibfnamefont {B.}~\bibnamefont {Woltjes}}, \bibinfo
  {author} {\bibfnamefont {E.~K.}\ \bibnamefont {Ko}}, \bibinfo {author}
  {\bibfnamefont {T.~W.}\ \bibnamefont {Noh}}, \bibinfo {author} {\bibfnamefont
  {W.}~\bibnamefont {Wu}}, \ and\ \bibinfo {author} {\bibfnamefont {J.~W.}\
  \bibnamefont {Robinson}},\ }\href@noop {} {\bibfield  {journal} {\bibinfo
  {journal} {Physical Review Materials}\ }\textbf {\bibinfo {volume} {4}},\
  \bibinfo {pages} {054414} (\bibinfo {year} {2020})}\BibitemShut {NoStop}%
\bibitem [{\citenamefont {Matsuno}\ \emph {et~al.}(2016)\citenamefont
  {Matsuno}, \citenamefont {Ogawa}, \citenamefont {Yasuda}, \citenamefont
  {Kagawa}, \citenamefont {Koshibae}, \citenamefont {Nagaosa}, \citenamefont
  {Tokura},\ and\ \citenamefont {Kawasaki}}]{matsuno2016interface}%
  \BibitemOpen
  \bibfield  {author} {\bibinfo {author} {\bibfnamefont {J.}~\bibnamefont
  {Matsuno}}, \bibinfo {author} {\bibfnamefont {N.}~\bibnamefont {Ogawa}},
  \bibinfo {author} {\bibfnamefont {K.}~\bibnamefont {Yasuda}}, \bibinfo
  {author} {\bibfnamefont {F.}~\bibnamefont {Kagawa}}, \bibinfo {author}
  {\bibfnamefont {W.}~\bibnamefont {Koshibae}}, \bibinfo {author}
  {\bibfnamefont {N.}~\bibnamefont {Nagaosa}}, \bibinfo {author} {\bibfnamefont
  {Y.}~\bibnamefont {Tokura}}, \ and\ \bibinfo {author} {\bibfnamefont
  {M.}~\bibnamefont {Kawasaki}},\ }\href@noop {} {\bibfield  {journal}
  {\bibinfo  {journal} {Science advances}\ }\textbf {\bibinfo {volume} {2}},\
  \bibinfo {pages} {e1600304} (\bibinfo {year} {2016})}\BibitemShut {NoStop}%
\bibitem [{\citenamefont {Ziese}\ \emph {et~al.}(2020)\citenamefont {Ziese},
  \citenamefont {Bern}, \citenamefont {Esquinazi},\ and\ \citenamefont
  {Lindfors-Vrejoiu}}]{ziese2020topological}%
  \BibitemOpen
  \bibfield  {author} {\bibinfo {author} {\bibfnamefont {M.}~\bibnamefont
  {Ziese}}, \bibinfo {author} {\bibfnamefont {F.}~\bibnamefont {Bern}},
  \bibinfo {author} {\bibfnamefont {P.~D.}\ \bibnamefont {Esquinazi}}, \ and\
  \bibinfo {author} {\bibfnamefont {I.}~\bibnamefont {Lindfors-Vrejoiu}},\
  }\href@noop {} {\bibfield  {journal} {\bibinfo  {journal} {physica status
  solidi (b)}\ }\textbf {\bibinfo {volume} {257}},\ \bibinfo {pages} {1900628}
  (\bibinfo {year} {2020})}\BibitemShut {NoStop}%
\bibitem [{\citenamefont {Xia}\ \emph {et~al.}(2009)\citenamefont {Xia},
  \citenamefont {Siemons}, \citenamefont {Koster}, \citenamefont {Beasley},\
  and\ \citenamefont {Kapitulnik}}]{xia2009critical}%
  \BibitemOpen
  \bibfield  {author} {\bibinfo {author} {\bibfnamefont {J.}~\bibnamefont
  {Xia}}, \bibinfo {author} {\bibfnamefont {W.}~\bibnamefont {Siemons}},
  \bibinfo {author} {\bibfnamefont {G.}~\bibnamefont {Koster}}, \bibinfo
  {author} {\bibfnamefont {M.}~\bibnamefont {Beasley}}, \ and\ \bibinfo
  {author} {\bibfnamefont {A.}~\bibnamefont {Kapitulnik}},\ }\href@noop {}
  {\bibfield  {journal} {\bibinfo  {journal} {Physical Review B}\ }\textbf
  {\bibinfo {volume} {79}},\ \bibinfo {pages} {140407} (\bibinfo {year}
  {2009})}\BibitemShut {NoStop}%
\bibitem [{\citenamefont {Chang}\ \emph {et~al.}(2009)\citenamefont {Chang},
  \citenamefont {Kim}, \citenamefont {Phark}, \citenamefont {Kim},
  \citenamefont {Yu},\ and\ \citenamefont {Noh}}]{chang2009fundamental}%
  \BibitemOpen
  \bibfield  {author} {\bibinfo {author} {\bibfnamefont {Y.~J.}\ \bibnamefont
  {Chang}}, \bibinfo {author} {\bibfnamefont {C.~H.}\ \bibnamefont {Kim}},
  \bibinfo {author} {\bibfnamefont {S.-H.}\ \bibnamefont {Phark}}, \bibinfo
  {author} {\bibfnamefont {Y.}~\bibnamefont {Kim}}, \bibinfo {author}
  {\bibfnamefont {J.}~\bibnamefont {Yu}}, \ and\ \bibinfo {author}
  {\bibfnamefont {T.}~\bibnamefont {Noh}},\ }\href@noop {} {\bibfield
  {journal} {\bibinfo  {journal} {Physical Review Letters}\ }\textbf {\bibinfo
  {volume} {103}},\ \bibinfo {pages} {057201} (\bibinfo {year}
  {2009})}\BibitemShut {NoStop}%
\bibitem [{\citenamefont {Ishigami}\ \emph {et~al.}(2015)\citenamefont
  {Ishigami}, \citenamefont {Yoshimatsu}, \citenamefont {Toyota}, \citenamefont
  {Takizawa}, \citenamefont {Yoshida}, \citenamefont {Shibata}, \citenamefont
  {Harano}, \citenamefont {Takahashi}, \citenamefont {Kadono}, \citenamefont
  {Verma} \emph {et~al.}}]{ishigami2015thickness}%
  \BibitemOpen
  \bibfield  {author} {\bibinfo {author} {\bibfnamefont {K.}~\bibnamefont
  {Ishigami}}, \bibinfo {author} {\bibfnamefont {K.}~\bibnamefont
  {Yoshimatsu}}, \bibinfo {author} {\bibfnamefont {D.}~\bibnamefont {Toyota}},
  \bibinfo {author} {\bibfnamefont {M.}~\bibnamefont {Takizawa}}, \bibinfo
  {author} {\bibfnamefont {T.}~\bibnamefont {Yoshida}}, \bibinfo {author}
  {\bibfnamefont {G.}~\bibnamefont {Shibata}}, \bibinfo {author} {\bibfnamefont
  {T.}~\bibnamefont {Harano}}, \bibinfo {author} {\bibfnamefont
  {Y.}~\bibnamefont {Takahashi}}, \bibinfo {author} {\bibfnamefont
  {T.}~\bibnamefont {Kadono}}, \bibinfo {author} {\bibfnamefont
  {V.}~\bibnamefont {Verma}},  \emph {et~al.},\ }\href@noop {} {\bibfield
  {journal} {\bibinfo  {journal} {Physical Review B}\ }\textbf {\bibinfo
  {volume} {92}},\ \bibinfo {pages} {064402} (\bibinfo {year}
  {2015})}\BibitemShut {NoStop}%
\bibitem [{\citenamefont {Schultz}\ \emph {et~al.}(2009)\citenamefont
  {Schultz}, \citenamefont {Reiner},\ and\ \citenamefont
  {Klein}}]{Schultz2009AHESROfilms}%
  \BibitemOpen
  \bibfield  {author} {\bibinfo {author} {\bibfnamefont {M.}~\bibnamefont
  {Schultz}}, \bibinfo {author} {\bibfnamefont {J.~W.}\ \bibnamefont {Reiner}},
  \ and\ \bibinfo {author} {\bibfnamefont {L.}~\bibnamefont {Klein}},\ }\href
  {\doibase 10.1063/1.3073935} {\bibfield  {journal} {\bibinfo  {journal}
  {Journal of Applied Physics}\ }\textbf {\bibinfo {volume} {105}},\ \bibinfo
  {pages} {07E906} (\bibinfo {year} {2009})}\BibitemShut {NoStop}%
\bibitem [{\citenamefont {Ziese}\ \emph {et~al.}(2010)\citenamefont {Ziese},
  \citenamefont {Vrejoiu},\ and\ \citenamefont
  {Hesse}}]{Ziese2010SROanisotropy}%
  \BibitemOpen
  \bibfield  {author} {\bibinfo {author} {\bibfnamefont {M.}~\bibnamefont
  {Ziese}}, \bibinfo {author} {\bibfnamefont {I.}~\bibnamefont {Vrejoiu}}, \
  and\ \bibinfo {author} {\bibfnamefont {D.}~\bibnamefont {Hesse}},\ }\href
  {\doibase 10.1103/PhysRevB.81.184418} {\bibfield  {journal} {\bibinfo
  {journal} {Phys. Rev. B}\ }\textbf {\bibinfo {volume} {81}},\ \bibinfo
  {pages} {184418} (\bibinfo {year} {2010})}\BibitemShut {NoStop}%
\bibitem [{\citenamefont {Wakabayashi}\ \emph {et~al.}(2021)\citenamefont
  {Wakabayashi}, \citenamefont {Kobayashi}, \citenamefont {Takeda},
  \citenamefont {Takiguchi}, \citenamefont {Irie}, \citenamefont {Fujimori},
  \citenamefont {Takeda}, \citenamefont {Okano}, \citenamefont {Krockenberger},
  \citenamefont {Taniyasu},\ and\ \citenamefont
  {Yamamoto}}]{Wakabayashi2021SROanisotropy}%
  \BibitemOpen
  \bibfield  {author} {\bibinfo {author} {\bibfnamefont {Y.~K.}\ \bibnamefont
  {Wakabayashi}}, \bibinfo {author} {\bibfnamefont {M.}~\bibnamefont
  {Kobayashi}}, \bibinfo {author} {\bibfnamefont {Y.}~\bibnamefont {Takeda}},
  \bibinfo {author} {\bibfnamefont {K.}~\bibnamefont {Takiguchi}}, \bibinfo
  {author} {\bibfnamefont {H.}~\bibnamefont {Irie}}, \bibinfo {author}
  {\bibfnamefont {S.-i.}\ \bibnamefont {Fujimori}}, \bibinfo {author}
  {\bibfnamefont {T.}~\bibnamefont {Takeda}}, \bibinfo {author} {\bibfnamefont
  {R.}~\bibnamefont {Okano}}, \bibinfo {author} {\bibfnamefont
  {Y.}~\bibnamefont {Krockenberger}}, \bibinfo {author} {\bibfnamefont
  {Y.}~\bibnamefont {Taniyasu}}, \ and\ \bibinfo {author} {\bibfnamefont
  {H.}~\bibnamefont {Yamamoto}},\ }\href {\doibase
  10.1103/PhysRevMaterials.5.124403} {\bibfield  {journal} {\bibinfo  {journal}
  {Phys. Rev. Materials}\ }\textbf {\bibinfo {volume} {5}},\ \bibinfo {pages}
  {124403} (\bibinfo {year} {2021})}\BibitemShut {NoStop}%
\bibitem [{\citenamefont {Ziese}\ and\ \citenamefont
  {Lindfors-Vrejoiu}(2018)}]{Ziese2018asymmetricSRO}%
  \BibitemOpen
  \bibfield  {author} {\bibinfo {author} {\bibfnamefont {M.}~\bibnamefont
  {Ziese}}\ and\ \bibinfo {author} {\bibfnamefont {I.}~\bibnamefont
  {Lindfors-Vrejoiu}},\ }\href {\doibase 10.1063/1.5051812} {\bibfield
  {journal} {\bibinfo  {journal} {Journal of Applied Physics}\ }\textbf
  {\bibinfo {volume} {124}},\ \bibinfo {pages} {163905} (\bibinfo {year}
  {2018})}\BibitemShut {NoStop}%
\bibitem [{\citenamefont {Sohn}\ \emph {et~al.}(2021)\citenamefont {Sohn},
  \citenamefont {Kim}, \citenamefont {Park}, \citenamefont {Choi},
  \citenamefont {Moon}, \citenamefont {Choi}, \citenamefont {Choi},
  \citenamefont {Zhou}, \citenamefont {Choi}, \citenamefont {Bombardi} \emph
  {et~al.}}]{sohn2021stable}%
  \BibitemOpen
  \bibfield  {author} {\bibinfo {author} {\bibfnamefont {B.}~\bibnamefont
  {Sohn}}, \bibinfo {author} {\bibfnamefont {B.}~\bibnamefont {Kim}}, \bibinfo
  {author} {\bibfnamefont {S.~Y.}\ \bibnamefont {Park}}, \bibinfo {author}
  {\bibfnamefont {H.~Y.}\ \bibnamefont {Choi}}, \bibinfo {author}
  {\bibfnamefont {J.~Y.}\ \bibnamefont {Moon}}, \bibinfo {author}
  {\bibfnamefont {T.}~\bibnamefont {Choi}}, \bibinfo {author} {\bibfnamefont
  {Y.~J.}\ \bibnamefont {Choi}}, \bibinfo {author} {\bibfnamefont
  {H.}~\bibnamefont {Zhou}}, \bibinfo {author} {\bibfnamefont {J.~W.}\
  \bibnamefont {Choi}}, \bibinfo {author} {\bibfnamefont {A.}~\bibnamefont
  {Bombardi}},  \emph {et~al.},\ }\href@noop {} {\bibfield  {journal} {\bibinfo
   {journal} {Physical Review Research}\ }\textbf {\bibinfo {volume} {3}},\
  \bibinfo {pages} {023232} (\bibinfo {year} {2021})}\BibitemShut {NoStop}%
\bibitem [{\citenamefont {Huang}\ \emph {et~al.}(2020)\citenamefont {Huang},
  \citenamefont {Lee}, \citenamefont {Kim}, \citenamefont {Sohn}, \citenamefont
  {Kim}, \citenamefont {Kao},\ and\ \citenamefont {Lee}}]{huang2020detection}%
  \BibitemOpen
  \bibfield  {author} {\bibinfo {author} {\bibfnamefont {H.}~\bibnamefont
  {Huang}}, \bibinfo {author} {\bibfnamefont {S.-J.}\ \bibnamefont {Lee}},
  \bibinfo {author} {\bibfnamefont {B.}~\bibnamefont {Kim}}, \bibinfo {author}
  {\bibfnamefont {B.}~\bibnamefont {Sohn}}, \bibinfo {author} {\bibfnamefont
  {C.}~\bibnamefont {Kim}}, \bibinfo {author} {\bibfnamefont {C.-C.}\
  \bibnamefont {Kao}}, \ and\ \bibinfo {author} {\bibfnamefont {J.-S.}\
  \bibnamefont {Lee}},\ }\href@noop {} {\bibfield  {journal} {\bibinfo
  {journal} {ACS Applied Materials \& Interfaces}\ }\textbf {\bibinfo {volume}
  {12}},\ \bibinfo {pages} {37757} (\bibinfo {year} {2020})}\BibitemShut
  {NoStop}%
\bibitem [{\citenamefont {Lu}\ \emph {et~al.}(2021)\citenamefont {Lu},
  \citenamefont {Si}, \citenamefont {Zhang}, \citenamefont {Tian},
  \citenamefont {Liu}, \citenamefont {Song}, \citenamefont {Dong},
  \citenamefont {Wang}, \citenamefont {Cheng}, \citenamefont {Qu} \emph
  {et~al.}}]{lu2021defect}%
  \BibitemOpen
  \bibfield  {author} {\bibinfo {author} {\bibfnamefont {J.}~\bibnamefont
  {Lu}}, \bibinfo {author} {\bibfnamefont {L.}~\bibnamefont {Si}}, \bibinfo
  {author} {\bibfnamefont {Q.}~\bibnamefont {Zhang}}, \bibinfo {author}
  {\bibfnamefont {C.}~\bibnamefont {Tian}}, \bibinfo {author} {\bibfnamefont
  {X.}~\bibnamefont {Liu}}, \bibinfo {author} {\bibfnamefont {C.}~\bibnamefont
  {Song}}, \bibinfo {author} {\bibfnamefont {S.}~\bibnamefont {Dong}}, \bibinfo
  {author} {\bibfnamefont {J.}~\bibnamefont {Wang}}, \bibinfo {author}
  {\bibfnamefont {S.}~\bibnamefont {Cheng}}, \bibinfo {author} {\bibfnamefont
  {L.}~\bibnamefont {Qu}},  \emph {et~al.},\ }\href@noop {} {\bibfield
  {journal} {\bibinfo  {journal} {Advanced Materials}\ }\textbf {\bibinfo
  {volume} {33}},\ \bibinfo {pages} {2102525} (\bibinfo {year}
  {2021})}\BibitemShut {NoStop}%
\bibitem [{\citenamefont {Wysocki}\ \emph {et~al.}(2020)\citenamefont
  {Wysocki}, \citenamefont {Yang}, \citenamefont {Gunkel}, \citenamefont
  {Dittmann}, \citenamefont {van Loosdrecht},\ and\ \citenamefont
  {Lindfors-Vrejoiu}}]{wysocki2020validity}%
  \BibitemOpen
  \bibfield  {author} {\bibinfo {author} {\bibfnamefont {L.}~\bibnamefont
  {Wysocki}}, \bibinfo {author} {\bibfnamefont {L.}~\bibnamefont {Yang}},
  \bibinfo {author} {\bibfnamefont {F.}~\bibnamefont {Gunkel}}, \bibinfo
  {author} {\bibfnamefont {R.}~\bibnamefont {Dittmann}}, \bibinfo {author}
  {\bibfnamefont {P.~H.}\ \bibnamefont {van Loosdrecht}}, \ and\ \bibinfo
  {author} {\bibfnamefont {I.}~\bibnamefont {Lindfors-Vrejoiu}},\ }\href@noop
  {} {\bibfield  {journal} {\bibinfo  {journal} {Physical review materials}\
  }\textbf {\bibinfo {volume} {4}},\ \bibinfo {pages} {054402} (\bibinfo {year}
  {2020})}\BibitemShut {NoStop}%
\bibitem [{\citenamefont {Ziese}\ \emph {et~al.}(2019)\citenamefont {Ziese},
  \citenamefont {Jin},\ and\ \citenamefont
  {Lindfors-Vrejoiu}}]{ziese2019unconventional}%
  \BibitemOpen
  \bibfield  {author} {\bibinfo {author} {\bibfnamefont {M.}~\bibnamefont
  {Ziese}}, \bibinfo {author} {\bibfnamefont {L.}~\bibnamefont {Jin}}, \ and\
  \bibinfo {author} {\bibfnamefont {I.}~\bibnamefont {Lindfors-Vrejoiu}},\
  }\href@noop {} {\bibfield  {journal} {\bibinfo  {journal} {Journal of
  Physics: Materials}\ }\textbf {\bibinfo {volume} {2}},\ \bibinfo {pages}
  {034008} (\bibinfo {year} {2019})}\BibitemShut {NoStop}%
\bibitem [{\citenamefont {Lu}\ \emph {et~al.}(2013)\citenamefont {Lu},
  \citenamefont {Yang}, \citenamefont {Song}, \citenamefont {Chow},\ and\
  \citenamefont {Chen}}]{lu2013control}%
  \BibitemOpen
  \bibfield  {author} {\bibinfo {author} {\bibfnamefont {W.}~\bibnamefont
  {Lu}}, \bibinfo {author} {\bibfnamefont {P.}~\bibnamefont {Yang}}, \bibinfo
  {author} {\bibfnamefont {W.~D.}\ \bibnamefont {Song}}, \bibinfo {author}
  {\bibfnamefont {G.~M.}\ \bibnamefont {Chow}}, \ and\ \bibinfo {author}
  {\bibfnamefont {J.~S.}\ \bibnamefont {Chen}},\ }\href@noop {} {\bibfield
  {journal} {\bibinfo  {journal} {Physical Review B}\ }\textbf {\bibinfo
  {volume} {88}},\ \bibinfo {pages} {214115} (\bibinfo {year}
  {2013})}\BibitemShut {NoStop}%
\bibitem [{\citenamefont {Samanta}\ \emph {et~al.}(2021)\citenamefont
  {Samanta}, \citenamefont {Ležaić}, \citenamefont {Blügel},\ and\
  \citenamefont {Mokrousov}}]{kartik2021srotheory}%
  \BibitemOpen
  \bibfield  {author} {\bibinfo {author} {\bibfnamefont {K.}~\bibnamefont
  {Samanta}}, \bibinfo {author} {\bibfnamefont {M.}~\bibnamefont {Ležaić}},
  \bibinfo {author} {\bibfnamefont {S.}~\bibnamefont {Blügel}}, \ and\
  \bibinfo {author} {\bibfnamefont {Y.}~\bibnamefont {Mokrousov}},\ }\href
  {\doibase 10.1063/5.0043742} {\bibfield  {journal} {\bibinfo  {journal}
  {Journal of Applied Physics}\ }\textbf {\bibinfo {volume} {129}},\ \bibinfo
  {pages} {093904} (\bibinfo {year} {2021})}\BibitemShut {NoStop}%
\bibitem [{\citenamefont {Kiyama}\ \emph {et~al.}(1996)\citenamefont {Kiyama},
  \citenamefont {Yoshimura}, \citenamefont {Kosuge}, \citenamefont {Ikeda},\
  and\ \citenamefont {Bando}}]{kiyama1996invar}%
  \BibitemOpen
  \bibfield  {author} {\bibinfo {author} {\bibfnamefont {T.}~\bibnamefont
  {Kiyama}}, \bibinfo {author} {\bibfnamefont {K.}~\bibnamefont {Yoshimura}},
  \bibinfo {author} {\bibfnamefont {K.}~\bibnamefont {Kosuge}}, \bibinfo
  {author} {\bibfnamefont {Y.}~\bibnamefont {Ikeda}}, \ and\ \bibinfo {author}
  {\bibfnamefont {Y.}~\bibnamefont {Bando}},\ }\href@noop {} {\bibfield
  {journal} {\bibinfo  {journal} {Physical Review B}\ }\textbf {\bibinfo
  {volume} {54}},\ \bibinfo {pages} {R756} (\bibinfo {year}
  {1996})}\BibitemShut {NoStop}%
\bibitem [{\citenamefont {Jones}\ \emph {et~al.}(1989)\citenamefont {Jones},
  \citenamefont {Battle}, \citenamefont {Lightfoot},\ and\ \citenamefont
  {Harrison}}]{jones1989structure}%
  \BibitemOpen
  \bibfield  {author} {\bibinfo {author} {\bibfnamefont {C.}~\bibnamefont
  {Jones}}, \bibinfo {author} {\bibfnamefont {P.}~\bibnamefont {Battle}},
  \bibinfo {author} {\bibfnamefont {P.}~\bibnamefont {Lightfoot}}, \ and\
  \bibinfo {author} {\bibfnamefont {W.}~\bibnamefont {Harrison}},\ }\href@noop
  {} {\bibfield  {journal} {\bibinfo  {journal} {Acta Crystallographica Section
  C: Crystal Structure Communications}\ }\textbf {\bibinfo {volume} {45}},\
  \bibinfo {pages} {365} (\bibinfo {year} {1989})}\BibitemShut {NoStop}%
\bibitem [{\citenamefont {Glazer}(1972)}]{glazer1972classification}%
  \BibitemOpen
  \bibfield  {author} {\bibinfo {author} {\bibfnamefont {A.}~\bibnamefont
  {Glazer}},\ }\href@noop {} {\bibfield  {journal} {\bibinfo  {journal} {Acta
  Crystallographica Section B: Structural Crystallography and Crystal
  Chemistry}\ }\textbf {\bibinfo {volume} {28}},\ \bibinfo {pages} {3384}
  (\bibinfo {year} {1972})}\BibitemShut {NoStop}%
\bibitem [{\citenamefont {Woodward}\ and\ \citenamefont
  {Reaney}(2005)}]{woodward2005electron}%
  \BibitemOpen
  \bibfield  {author} {\bibinfo {author} {\bibfnamefont {D.~I.}\ \bibnamefont
  {Woodward}}\ and\ \bibinfo {author} {\bibfnamefont {I.~M.}\ \bibnamefont
  {Reaney}},\ }\href@noop {} {\bibfield  {journal} {\bibinfo  {journal} {Acta
  Crystallographica Section B: Structural Science}\ }\textbf {\bibinfo {volume}
  {61}},\ \bibinfo {pages} {387} (\bibinfo {year} {2005})}\BibitemShut
  {NoStop}%
\bibitem [{\citenamefont {Vailionis}\ \emph {et~al.}(2008)\citenamefont
  {Vailionis}, \citenamefont {Siemons},\ and\ \citenamefont
  {Koster}}]{vailionis2008room}%
  \BibitemOpen
  \bibfield  {author} {\bibinfo {author} {\bibfnamefont {A.}~\bibnamefont
  {Vailionis}}, \bibinfo {author} {\bibfnamefont {W.}~\bibnamefont {Siemons}},
  \ and\ \bibinfo {author} {\bibfnamefont {G.}~\bibnamefont {Koster}},\
  }\href@noop {} {\bibfield  {journal} {\bibinfo  {journal} {Applied physics
  letters}\ }\textbf {\bibinfo {volume} {93}},\ \bibinfo {pages} {051909}
  (\bibinfo {year} {2008})}\BibitemShut {NoStop}%
\bibitem [{\citenamefont {Gao}\ \emph {et~al.}(2016)\citenamefont {Gao},
  \citenamefont {Dong}, \citenamefont {Xu}, \citenamefont {Zhou}, \citenamefont
  {Yuan}, \citenamefont {Gopalan}, \citenamefont {Gao}, \citenamefont {Fong},
  \citenamefont {Chen}, \citenamefont {Luo} \emph
  {et~al.}}]{gao2016interfacial}%
  \BibitemOpen
  \bibfield  {author} {\bibinfo {author} {\bibfnamefont {R.}~\bibnamefont
  {Gao}}, \bibinfo {author} {\bibfnamefont {Y.}~\bibnamefont {Dong}}, \bibinfo
  {author} {\bibfnamefont {H.}~\bibnamefont {Xu}}, \bibinfo {author}
  {\bibfnamefont {H.}~\bibnamefont {Zhou}}, \bibinfo {author} {\bibfnamefont
  {Y.}~\bibnamefont {Yuan}}, \bibinfo {author} {\bibfnamefont {V.}~\bibnamefont
  {Gopalan}}, \bibinfo {author} {\bibfnamefont {C.}~\bibnamefont {Gao}},
  \bibinfo {author} {\bibfnamefont {D.~D.}\ \bibnamefont {Fong}}, \bibinfo
  {author} {\bibfnamefont {Z.}~\bibnamefont {Chen}}, \bibinfo {author}
  {\bibfnamefont {Z.}~\bibnamefont {Luo}},  \emph {et~al.},\ }\href@noop {}
  {\bibfield  {journal} {\bibinfo  {journal} {ACS applied materials \&
  interfaces}\ }\textbf {\bibinfo {volume} {8}},\ \bibinfo {pages} {14871}
  (\bibinfo {year} {2016})}\BibitemShut {NoStop}%
\bibitem [{\citenamefont {Aso}\ \emph {et~al.}(2013)\citenamefont {Aso},
  \citenamefont {Kan}, \citenamefont {Shimakawa},\ and\ \citenamefont
  {Kurata}}]{aso2013atomic}%
  \BibitemOpen
  \bibfield  {author} {\bibinfo {author} {\bibfnamefont {R.}~\bibnamefont
  {Aso}}, \bibinfo {author} {\bibfnamefont {D.}~\bibnamefont {Kan}}, \bibinfo
  {author} {\bibfnamefont {Y.}~\bibnamefont {Shimakawa}}, \ and\ \bibinfo
  {author} {\bibfnamefont {H.}~\bibnamefont {Kurata}},\ }\href@noop {}
  {\bibfield  {journal} {\bibinfo  {journal} {Scientific reports}\ }\textbf
  {\bibinfo {volume} {3}},\ \bibinfo {pages} {1} (\bibinfo {year}
  {2013})}\BibitemShut {NoStop}%
\bibitem [{\citenamefont {Chang}\ \emph {et~al.}(2011)\citenamefont {Chang},
  \citenamefont {Chang}, \citenamefont {Jang}, \citenamefont {Jeong},
  \citenamefont {Jung}, \citenamefont {Kim}, \citenamefont {Chung},\ and\
  \citenamefont {Noh}}]{chang2011thickness}%
  \BibitemOpen
  \bibfield  {author} {\bibinfo {author} {\bibfnamefont {S.~H.}\ \bibnamefont
  {Chang}}, \bibinfo {author} {\bibfnamefont {Y.~J.}\ \bibnamefont {Chang}},
  \bibinfo {author} {\bibfnamefont {S.}~\bibnamefont {Jang}}, \bibinfo {author}
  {\bibfnamefont {D.}~\bibnamefont {Jeong}}, \bibinfo {author} {\bibfnamefont
  {C.}~\bibnamefont {Jung}}, \bibinfo {author} {\bibfnamefont {Y.-J.}\
  \bibnamefont {Kim}}, \bibinfo {author} {\bibfnamefont {J.-S.}\ \bibnamefont
  {Chung}}, \ and\ \bibinfo {author} {\bibfnamefont {T.}~\bibnamefont {Noh}},\
  }\href@noop {} {\bibfield  {journal} {\bibinfo  {journal} {Physical Review
  B}\ }\textbf {\bibinfo {volume} {84}},\ \bibinfo {pages} {104101} (\bibinfo
  {year} {2011})}\BibitemShut {NoStop}%
\bibitem [{\citenamefont {Vailionis}\ \emph {et~al.}(2011)\citenamefont
  {Vailionis}, \citenamefont {Boschker}, \citenamefont {Siemons}, \citenamefont
  {Houwman}, \citenamefont {Blank}, \citenamefont {Rijnders},\ and\
  \citenamefont {Koster}}]{vailionis2011misfit}%
  \BibitemOpen
  \bibfield  {author} {\bibinfo {author} {\bibfnamefont {A.}~\bibnamefont
  {Vailionis}}, \bibinfo {author} {\bibfnamefont {H.}~\bibnamefont {Boschker}},
  \bibinfo {author} {\bibfnamefont {W.}~\bibnamefont {Siemons}}, \bibinfo
  {author} {\bibfnamefont {E.~P.}\ \bibnamefont {Houwman}}, \bibinfo {author}
  {\bibfnamefont {D.~H.}\ \bibnamefont {Blank}}, \bibinfo {author}
  {\bibfnamefont {G.}~\bibnamefont {Rijnders}}, \ and\ \bibinfo {author}
  {\bibfnamefont {G.}~\bibnamefont {Koster}},\ }\href@noop {} {\bibfield
  {journal} {\bibinfo  {journal} {Physical Review B}\ }\textbf {\bibinfo
  {volume} {83}},\ \bibinfo {pages} {064101} (\bibinfo {year}
  {2011})}\BibitemShut {NoStop}%
\bibitem [{\citenamefont {Maria}\ \emph {et~al.}(2000)\citenamefont {Maria},
  \citenamefont {McKinstry},\ and\ \citenamefont
  {Trolier-McKinstry}}]{maria2000origin}%
  \BibitemOpen
  \bibfield  {author} {\bibinfo {author} {\bibfnamefont {J.-P.}\ \bibnamefont
  {Maria}}, \bibinfo {author} {\bibfnamefont {H.}~\bibnamefont {McKinstry}}, \
  and\ \bibinfo {author} {\bibfnamefont {S.}~\bibnamefont
  {Trolier-McKinstry}},\ }\href@noop {} {\bibfield  {journal} {\bibinfo
  {journal} {Applied Physics Letters}\ }\textbf {\bibinfo {volume} {76}},\
  \bibinfo {pages} {3382} (\bibinfo {year} {2000})}\BibitemShut {NoStop}%
\bibitem [{\citenamefont {Rao}\ \emph {et~al.}(1997)\citenamefont {Rao},
  \citenamefont {Gan},\ and\ \citenamefont {Eom}}]{rao1997growth}%
  \BibitemOpen
  \bibfield  {author} {\bibinfo {author} {\bibfnamefont {R.~A.}\ \bibnamefont
  {Rao}}, \bibinfo {author} {\bibfnamefont {Q.}~\bibnamefont {Gan}}, \ and\
  \bibinfo {author} {\bibfnamefont {C.-B.}\ \bibnamefont {Eom}},\ }\href@noop
  {} {\bibfield  {journal} {\bibinfo  {journal} {Applied physics letters}\
  }\textbf {\bibinfo {volume} {71}},\ \bibinfo {pages} {1171} (\bibinfo {year}
  {1997})}\BibitemShut {NoStop}%
\bibitem [{\citenamefont {Gan}\ \emph {et~al.}(1998)\citenamefont {Gan},
  \citenamefont {Rao}, \citenamefont {Eom}, \citenamefont {Garrett},\ and\
  \citenamefont {Lee}}]{gan1998direct}%
  \BibitemOpen
  \bibfield  {author} {\bibinfo {author} {\bibfnamefont {Q.}~\bibnamefont
  {Gan}}, \bibinfo {author} {\bibfnamefont {R.}~\bibnamefont {Rao}}, \bibinfo
  {author} {\bibfnamefont {C.}~\bibnamefont {Eom}}, \bibinfo {author}
  {\bibfnamefont {J.}~\bibnamefont {Garrett}}, \ and\ \bibinfo {author}
  {\bibfnamefont {M.}~\bibnamefont {Lee}},\ }\href@noop {} {\bibfield
  {journal} {\bibinfo  {journal} {Applied Physics Letters}\ }\textbf {\bibinfo
  {volume} {72}},\ \bibinfo {pages} {978} (\bibinfo {year} {1998})}\BibitemShut
  {NoStop}%
\bibitem [{\citenamefont {Schlep{\"u}tz}\ \emph {et~al.}(2005)\citenamefont
  {Schlep{\"u}tz}, \citenamefont {Herger}, \citenamefont {Willmott},
  \citenamefont {Patterson}, \citenamefont {Bunk}, \citenamefont
  {Br{\"o}nnimann}, \citenamefont {Henrich}, \citenamefont {H{\"u}lsen},\ and\
  \citenamefont {Eikenberry}}]{schleputz2005improved}%
  \BibitemOpen
  \bibfield  {author} {\bibinfo {author} {\bibfnamefont {C.}~\bibnamefont
  {Schlep{\"u}tz}}, \bibinfo {author} {\bibfnamefont {R.}~\bibnamefont
  {Herger}}, \bibinfo {author} {\bibfnamefont {P.}~\bibnamefont {Willmott}},
  \bibinfo {author} {\bibfnamefont {B.}~\bibnamefont {Patterson}}, \bibinfo
  {author} {\bibfnamefont {O.}~\bibnamefont {Bunk}}, \bibinfo {author}
  {\bibfnamefont {C.}~\bibnamefont {Br{\"o}nnimann}}, \bibinfo {author}
  {\bibfnamefont {B.}~\bibnamefont {Henrich}}, \bibinfo {author} {\bibfnamefont
  {G.}~\bibnamefont {H{\"u}lsen}}, \ and\ \bibinfo {author} {\bibfnamefont
  {E.}~\bibnamefont {Eikenberry}},\ }\href@noop {} {\bibfield  {journal}
  {\bibinfo  {journal} {Acta Crystallographica Section A: Foundations of
  Crystallography}\ }\textbf {\bibinfo {volume} {61}},\ \bibinfo {pages} {418}
  (\bibinfo {year} {2005})}\BibitemShut {NoStop}%
\bibitem [{\citenamefont {Bj{\"o}rck}\ and\ \citenamefont
  {Andersson}(2007)}]{bjorck2007genx}%
  \BibitemOpen
  \bibfield  {author} {\bibinfo {author} {\bibfnamefont {M.}~\bibnamefont
  {Bj{\"o}rck}}\ and\ \bibinfo {author} {\bibfnamefont {G.}~\bibnamefont
  {Andersson}},\ }\href@noop {} {\bibfield  {journal} {\bibinfo  {journal}
  {Journal of Applied Crystallography}\ }\textbf {\bibinfo {volume} {40}},\
  \bibinfo {pages} {1174} (\bibinfo {year} {2007})}\BibitemShut {NoStop}%
\bibitem [{\citenamefont {Koohfar}\ \emph {et~al.}(2019)\citenamefont
  {Koohfar}, \citenamefont {Georgescu}, \citenamefont {Penn}, \citenamefont
  {LeBeau}, \citenamefont {Arenholz},\ and\ \citenamefont
  {Kumah}}]{koohfar2017confinement}%
  \BibitemOpen
  \bibfield  {author} {\bibinfo {author} {\bibfnamefont {S.}~\bibnamefont
  {Koohfar}}, \bibinfo {author} {\bibfnamefont {A.~B.}\ \bibnamefont
  {Georgescu}}, \bibinfo {author} {\bibfnamefont {A.~N.}\ \bibnamefont {Penn}},
  \bibinfo {author} {\bibfnamefont {J.~M.}\ \bibnamefont {LeBeau}}, \bibinfo
  {author} {\bibfnamefont {E.}~\bibnamefont {Arenholz}}, \ and\ \bibinfo
  {author} {\bibfnamefont {D.~P.}\ \bibnamefont {Kumah}},\ }\href@noop {}
  {\bibfield  {journal} {\bibinfo  {journal} {npj Quantum Materials}\ }\textbf
  {\bibinfo {volume} {4}},\ \bibinfo {pages} {1} (\bibinfo {year}
  {2019})}\BibitemShut {NoStop}%
\bibitem [{\citenamefont {Sang}\ and\ \citenamefont
  {LeBeau}(2014)}]{sang2014revolving}%
  \BibitemOpen
  \bibfield  {author} {\bibinfo {author} {\bibfnamefont {X.}~\bibnamefont
  {Sang}}\ and\ \bibinfo {author} {\bibfnamefont {J.~M.}\ \bibnamefont
  {LeBeau}},\ }\href@noop {} {\bibfield  {journal} {\bibinfo  {journal}
  {Ultramicroscopy}\ }\textbf {\bibinfo {volume} {138}},\ \bibinfo {pages} {28}
  (\bibinfo {year} {2014})}\BibitemShut {NoStop}%
\bibitem [{\citenamefont {May}\ \emph {et~al.}(2010)\citenamefont {May},
  \citenamefont {Kim}, \citenamefont {Rondinelli}, \citenamefont {Karapetrova},
  \citenamefont {Spaldin}, \citenamefont {Bhattacharya},\ and\ \citenamefont
  {Ryan}}]{May2010LaNiO3}%
  \BibitemOpen
  \bibfield  {author} {\bibinfo {author} {\bibfnamefont {S.~J.}\ \bibnamefont
  {May}}, \bibinfo {author} {\bibfnamefont {J.-W.}\ \bibnamefont {Kim}},
  \bibinfo {author} {\bibfnamefont {J.~M.}\ \bibnamefont {Rondinelli}},
  \bibinfo {author} {\bibfnamefont {E.}~\bibnamefont {Karapetrova}}, \bibinfo
  {author} {\bibfnamefont {N.~A.}\ \bibnamefont {Spaldin}}, \bibinfo {author}
  {\bibfnamefont {A.}~\bibnamefont {Bhattacharya}}, \ and\ \bibinfo {author}
  {\bibfnamefont {P.~J.}\ \bibnamefont {Ryan}},\ }\href {\doibase
  10.1103/PhysRevB.82.014110} {\bibfield  {journal} {\bibinfo  {journal} {Phys.
  Rev. B}\ }\textbf {\bibinfo {volume} {82}},\ \bibinfo {pages} {014110}
  (\bibinfo {year} {2010})}\BibitemShut {NoStop}%
\bibitem [{\citenamefont {Siwakoti}\ \emph {et~al.}(2021)\citenamefont
  {Siwakoti}, \citenamefont {Wang}, \citenamefont {Saghayezhian}, \citenamefont
  {Howe}, \citenamefont {Ali}, \citenamefont {Zhu},\ and\ \citenamefont
  {Zhang}}]{siwakoti2021abrupt}%
  \BibitemOpen
  \bibfield  {author} {\bibinfo {author} {\bibfnamefont {P.}~\bibnamefont
  {Siwakoti}}, \bibinfo {author} {\bibfnamefont {Z.}~\bibnamefont {Wang}},
  \bibinfo {author} {\bibfnamefont {M.}~\bibnamefont {Saghayezhian}}, \bibinfo
  {author} {\bibfnamefont {D.}~\bibnamefont {Howe}}, \bibinfo {author}
  {\bibfnamefont {Z.}~\bibnamefont {Ali}}, \bibinfo {author} {\bibfnamefont
  {Y.}~\bibnamefont {Zhu}}, \ and\ \bibinfo {author} {\bibfnamefont
  {J.}~\bibnamefont {Zhang}},\ }\href {\doibase
  10.1103/PhysRevMaterials.5.114409} {\bibfield  {journal} {\bibinfo  {journal}
  {Phys. Rev. Materials}\ }\textbf {\bibinfo {volume} {5}},\ \bibinfo {pages}
  {114409} (\bibinfo {year} {2021})}\BibitemShut {NoStop}%
\bibitem [{\citenamefont {Wang}\ \emph {et~al.}(2020)\citenamefont {Wang},
  \citenamefont {Li}, \citenamefont {Liu}, \citenamefont {Chen}, \citenamefont
  {Ji}, \citenamefont {Wang}, \citenamefont {Cheng}, \citenamefont {Lu},
  \citenamefont {Rijnders}, \citenamefont {Koster} \emph
  {et~al.}}]{wang2020magnetic}%
  \BibitemOpen
  \bibfield  {author} {\bibinfo {author} {\bibfnamefont {W.}~\bibnamefont
  {Wang}}, \bibinfo {author} {\bibfnamefont {L.}~\bibnamefont {Li}}, \bibinfo
  {author} {\bibfnamefont {J.}~\bibnamefont {Liu}}, \bibinfo {author}
  {\bibfnamefont {B.}~\bibnamefont {Chen}}, \bibinfo {author} {\bibfnamefont
  {Y.}~\bibnamefont {Ji}}, \bibinfo {author} {\bibfnamefont {J.}~\bibnamefont
  {Wang}}, \bibinfo {author} {\bibfnamefont {G.}~\bibnamefont {Cheng}},
  \bibinfo {author} {\bibfnamefont {Y.}~\bibnamefont {Lu}}, \bibinfo {author}
  {\bibfnamefont {G.}~\bibnamefont {Rijnders}}, \bibinfo {author}
  {\bibfnamefont {G.}~\bibnamefont {Koster}},  \emph {et~al.},\ }\href@noop {}
  {\bibfield  {journal} {\bibinfo  {journal} {npj Quantum Materials}\ }\textbf
  {\bibinfo {volume} {5}},\ \bibinfo {pages} {1} (\bibinfo {year}
  {2020})}\BibitemShut {NoStop}%
\bibitem [{\citenamefont {Kar}\ \emph {et~al.}(2021)\citenamefont {Kar},
  \citenamefont {Singh}, \citenamefont {Yang}, \citenamefont {Lin},
  \citenamefont {Das}, \citenamefont {Hsu},\ and\ \citenamefont
  {Lee}}]{kar2021high}%
  \BibitemOpen
  \bibfield  {author} {\bibinfo {author} {\bibfnamefont {U.}~\bibnamefont
  {Kar}}, \bibinfo {author} {\bibfnamefont {A.~K.}\ \bibnamefont {Singh}},
  \bibinfo {author} {\bibfnamefont {S.}~\bibnamefont {Yang}}, \bibinfo {author}
  {\bibfnamefont {C.-Y.}\ \bibnamefont {Lin}}, \bibinfo {author} {\bibfnamefont
  {B.}~\bibnamefont {Das}}, \bibinfo {author} {\bibfnamefont {C.-H.}\
  \bibnamefont {Hsu}}, \ and\ \bibinfo {author} {\bibfnamefont {W.-L.}\
  \bibnamefont {Lee}},\ }\href@noop {} {\bibfield  {journal} {\bibinfo
  {journal} {Sci. Rep.}\ }\textbf {\bibinfo {volume} {11}},\ \bibinfo {pages}
  {16070} (\bibinfo {year} {2021})}\BibitemShut {NoStop}%
\bibitem [{\citenamefont {El-Mellouhi}\ \emph {et~al.}(2011)\citenamefont
  {El-Mellouhi}, \citenamefont {Brothers}, \citenamefont {Lucero},\ and\
  \citenamefont {Scuseria}}]{el2011modeling}%
  \BibitemOpen
  \bibfield  {author} {\bibinfo {author} {\bibfnamefont {F.}~\bibnamefont
  {El-Mellouhi}}, \bibinfo {author} {\bibfnamefont {E.~N.}\ \bibnamefont
  {Brothers}}, \bibinfo {author} {\bibfnamefont {M.~J.}\ \bibnamefont
  {Lucero}}, \ and\ \bibinfo {author} {\bibfnamefont {G.~E.}\ \bibnamefont
  {Scuseria}},\ }\href@noop {} {\bibfield  {journal} {\bibinfo  {journal}
  {Physical Review B}\ }\textbf {\bibinfo {volume} {84}},\ \bibinfo {pages}
  {115122} (\bibinfo {year} {2011})}\BibitemShut {NoStop}%
\bibitem [{\citenamefont {He}\ \emph {et~al.}(2010)\citenamefont {He},
  \citenamefont {Borisevich}, \citenamefont {Kalinin}, \citenamefont
  {Pennycook},\ and\ \citenamefont {Pantelides}}]{he2010control}%
  \BibitemOpen
  \bibfield  {author} {\bibinfo {author} {\bibfnamefont {J.}~\bibnamefont
  {He}}, \bibinfo {author} {\bibfnamefont {A.}~\bibnamefont {Borisevich}},
  \bibinfo {author} {\bibfnamefont {S.~V.}\ \bibnamefont {Kalinin}}, \bibinfo
  {author} {\bibfnamefont {S.~J.}\ \bibnamefont {Pennycook}}, \ and\ \bibinfo
  {author} {\bibfnamefont {S.~T.}\ \bibnamefont {Pantelides}},\ }\href@noop {}
  {\bibfield  {journal} {\bibinfo  {journal} {Physical review letters}\
  }\textbf {\bibinfo {volume} {105}},\ \bibinfo {pages} {227203} (\bibinfo
  {year} {2010})}\BibitemShut {NoStop}%
\bibitem [{\citenamefont {Dabrowski}\ \emph {et~al.}(2006)\citenamefont
  {Dabrowski}, \citenamefont {Kolesnik}, \citenamefont {Chmaissem},
  \citenamefont {Maxwell}, \citenamefont {Mais},\ and\ \citenamefont
  {Jorgensen}}]{dabrowski2006magnetic}%
  \BibitemOpen
  \bibfield  {author} {\bibinfo {author} {\bibfnamefont {B.}~\bibnamefont
  {Dabrowski}}, \bibinfo {author} {\bibfnamefont {S.}~\bibnamefont {Kolesnik}},
  \bibinfo {author} {\bibfnamefont {O.}~\bibnamefont {Chmaissem}}, \bibinfo
  {author} {\bibfnamefont {T.}~\bibnamefont {Maxwell}}, \bibinfo {author}
  {\bibfnamefont {J.}~\bibnamefont {Mais}}, \ and\ \bibinfo {author}
  {\bibfnamefont {J.}~\bibnamefont {Jorgensen}},\ }\href@noop {} {\bibfield
  {journal} {\bibinfo  {journal} {physica status solidi (b)}\ }\textbf
  {\bibinfo {volume} {243}},\ \bibinfo {pages} {13} (\bibinfo {year}
  {2006})}\BibitemShut {NoStop}%
\bibitem [{\citenamefont {Bushmeleva}\ \emph {et~al.}(2006)\citenamefont
  {Bushmeleva}, \citenamefont {Pomjakushin}, \citenamefont {Pomjakushina},
  \citenamefont {Sheptyakov},\ and\ \citenamefont
  {Balagurov}}]{bushmeleva2006evidence}%
  \BibitemOpen
  \bibfield  {author} {\bibinfo {author} {\bibfnamefont {S.}~\bibnamefont
  {Bushmeleva}}, \bibinfo {author} {\bibfnamefont {V.~Y.}\ \bibnamefont
  {Pomjakushin}}, \bibinfo {author} {\bibfnamefont {E.}~\bibnamefont
  {Pomjakushina}}, \bibinfo {author} {\bibfnamefont {D.}~\bibnamefont
  {Sheptyakov}}, \ and\ \bibinfo {author} {\bibfnamefont {A.}~\bibnamefont
  {Balagurov}},\ }\href@noop {} {\bibfield  {journal} {\bibinfo  {journal}
  {Journal of magnetism and magnetic materials}\ }\textbf {\bibinfo {volume}
  {305}},\ \bibinfo {pages} {491} (\bibinfo {year} {2006})}\BibitemShut
  {NoStop}%
\bibitem [{\citenamefont {Kan}\ \emph {et~al.}(2013)\citenamefont {Kan},
  \citenamefont {Aso}, \citenamefont {Kurata},\ and\ \citenamefont
  {Shimakawa}}]{Kan2013tetragonalSRO}%
  \BibitemOpen
  \bibfield  {author} {\bibinfo {author} {\bibfnamefont {D.}~\bibnamefont
  {Kan}}, \bibinfo {author} {\bibfnamefont {R.}~\bibnamefont {Aso}}, \bibinfo
  {author} {\bibfnamefont {H.}~\bibnamefont {Kurata}}, \ and\ \bibinfo {author}
  {\bibfnamefont {Y.}~\bibnamefont {Shimakawa}},\ }\href {\doibase
  10.1063/1.4803869} {\bibfield  {journal} {\bibinfo  {journal} {Journal of
  Applied Physics}\ }\textbf {\bibinfo {volume} {113}},\ \bibinfo {pages}
  {173912} (\bibinfo {year} {2013})}\BibitemShut {NoStop}%
\bibitem [{\citenamefont {Bern}\ \emph {et~al.}(2013)\citenamefont {Bern},
  \citenamefont {Ziese}, \citenamefont {Dörr}, \citenamefont {Herklotz},\ and\
  \citenamefont {Vrejoiu}}]{Bern2013AHE}%
  \BibitemOpen
  \bibfield  {author} {\bibinfo {author} {\bibfnamefont {F.}~\bibnamefont
  {Bern}}, \bibinfo {author} {\bibfnamefont {M.}~\bibnamefont {Ziese}},
  \bibinfo {author} {\bibfnamefont {K.}~\bibnamefont {Dörr}}, \bibinfo
  {author} {\bibfnamefont {A.}~\bibnamefont {Herklotz}}, \ and\ \bibinfo
  {author} {\bibfnamefont {I.}~\bibnamefont {Vrejoiu}},\ }\href@noop {}
  {\bibfield  {journal} {\bibinfo  {journal} {physica status solidi (RRL) –
  Rapid Research Letters}\ }\textbf {\bibinfo {volume} {7}},\ \bibinfo {pages}
  {204} (\bibinfo {year} {2013})}\BibitemShut {NoStop}%
\bibitem [{\citenamefont {Malsch}\ \emph {et~al.}(2020)\citenamefont {Malsch},
  \citenamefont {Ivaneyko}, \citenamefont {Milde}, \citenamefont {Wysocki},
  \citenamefont {Yang}, \citenamefont {van Loosdrecht}, \citenamefont
  {Lindfors-Vrejoiu},\ and\ \citenamefont {Eng}}]{malsch2020correlating}%
  \BibitemOpen
  \bibfield  {author} {\bibinfo {author} {\bibfnamefont {G.}~\bibnamefont
  {Malsch}}, \bibinfo {author} {\bibfnamefont {D.}~\bibnamefont {Ivaneyko}},
  \bibinfo {author} {\bibfnamefont {P.}~\bibnamefont {Milde}}, \bibinfo
  {author} {\bibfnamefont {L.}~\bibnamefont {Wysocki}}, \bibinfo {author}
  {\bibfnamefont {L.}~\bibnamefont {Yang}}, \bibinfo {author} {\bibfnamefont
  {P.~H.}\ \bibnamefont {van Loosdrecht}}, \bibinfo {author} {\bibfnamefont
  {I.}~\bibnamefont {Lindfors-Vrejoiu}}, \ and\ \bibinfo {author}
  {\bibfnamefont {L.~M.}\ \bibnamefont {Eng}},\ }\href@noop {} {\bibfield
  {journal} {\bibinfo  {journal} {ACS Applied Nano Materials}\ }\textbf
  {\bibinfo {volume} {3}},\ \bibinfo {pages} {1182} (\bibinfo {year}
  {2020})}\BibitemShut {NoStop}%
\bibitem [{\citenamefont {Caspi}\ \emph {et~al.}(2022)\citenamefont {Caspi},
  \citenamefont {Shoham}, \citenamefont {Baskin}, \citenamefont {Weinfeld},
  \citenamefont {Piamonteze}, \citenamefont {Stoerzinger},\ and\ \citenamefont
  {Kornblum}}]{Caspi2022SrVO3}%
  \BibitemOpen
  \bibfield  {author} {\bibinfo {author} {\bibfnamefont {S.}~\bibnamefont
  {Caspi}}, \bibinfo {author} {\bibfnamefont {L.}~\bibnamefont {Shoham}},
  \bibinfo {author} {\bibfnamefont {M.}~\bibnamefont {Baskin}}, \bibinfo
  {author} {\bibfnamefont {K.}~\bibnamefont {Weinfeld}}, \bibinfo {author}
  {\bibfnamefont {C.}~\bibnamefont {Piamonteze}}, \bibinfo {author}
  {\bibfnamefont {K.~A.}\ \bibnamefont {Stoerzinger}}, \ and\ \bibinfo {author}
  {\bibfnamefont {L.}~\bibnamefont {Kornblum}},\ }\href {\doibase
  10.1116/6.0001419} {\bibfield  {journal} {\bibinfo  {journal} {Journal of
  Vacuum Science \& Technology A}\ }\textbf {\bibinfo {volume} {40}},\ \bibinfo
  {pages} {013208} (\bibinfo {year} {2022})}\BibitemShut {NoStop}%
\bibitem [{\citenamefont {Shin}\ \emph {et~al.}(2005)\citenamefont {Shin},
  \citenamefont {Kalinin}, \citenamefont {Lee}, \citenamefont {Christen},
  \citenamefont {Moore}, \citenamefont {Plummer},\ and\ \citenamefont
  {Baddorf}}]{SHIN2005SROsurface-stability}%
  \BibitemOpen
  \bibfield  {author} {\bibinfo {author} {\bibfnamefont {J.}~\bibnamefont
  {Shin}}, \bibinfo {author} {\bibfnamefont {S.}~\bibnamefont {Kalinin}},
  \bibinfo {author} {\bibfnamefont {H.}~\bibnamefont {Lee}}, \bibinfo {author}
  {\bibfnamefont {H.}~\bibnamefont {Christen}}, \bibinfo {author}
  {\bibfnamefont {R.}~\bibnamefont {Moore}}, \bibinfo {author} {\bibfnamefont
  {E.}~\bibnamefont {Plummer}}, \ and\ \bibinfo {author} {\bibfnamefont
  {A.}~\bibnamefont {Baddorf}},\ }\href {\doibase
  https://doi.org/10.1016/j.susc.2005.02.038} {\bibfield  {journal} {\bibinfo
  {journal} {Surface Science}\ }\textbf {\bibinfo {volume} {581}},\ \bibinfo
  {pages} {118} (\bibinfo {year} {2005})}\BibitemShut {NoStop}%
\end{thebibliography}
\end{document}